\documentclass[fleqn,usenatbib]{mnras}

\usepackage{newtxtext,newtxmath}

\usepackage[T1]{fontenc}
\usepackage[dvipsnames]{xcolor}

\DeclareRobustCommand{\VAN}[3]{#2}
\let\VANthebibliography\thebibliography
\def\thebibliography{\DeclareRobustCommand{\VAN}[3]{##3}\VANthebibliography}

\usepackage{graphicx}	
\usepackage{amsmath}	
\usepackage{multirow}
\usepackage{booktabs}
\usepackage{calc}
\usepackage{xcolor}
\usepackage{adjustbox}
\usepackage{soul}

\title[Post-FUor Tracers]{Observational Chemical Signatures of the Past FU~Ori Outbursts}

\author[L. Zwicky et al.]{
Lis Zwicky,$^{1}$\thanks{E-mail: tsvikki.lis@urfu.ru}
Tamara Molyarova,$^{2,3}$
Vitaly Akimkin,$^{2}$
Grigorii V. Smirnov-Pinchukov,$^{4}$
\newauthor{
Dmitry Semenov$^{4,8}$, \'Agnes K\'osp\'al$^{5, 6, 7, 4}$, P\'eter \'Abrah\'am$^{5, 6, 7}$}
\\
$^{1}$Institute of Natural Sciences and Mathematics, Ural Federal University, 19 Mira Str., 620075, Ekaterinburg, Russia\\
$^{2}$Institute of Astronomy, Russian Academy of Sciences, 48 Pyatnitskaya St., Moscow, 119017, Russia\\
$^{3}${
Research Institute of Physics, Southern Federal University, Rostov-on-Don 344090, Russia}\\
$^{4}$Max Planck Institute for Astronomy, Königstuhl 17, D-69117 Heidelberg, Germany\\
$^{5}${Konkoly Observatory, HUN-REN Research Centre for Astronomy and Earth Sciences, Konkoly-Thege Mikl\'os \'ut 15-17, 1121 Budapest, Hungary}\\
$^{6}$CSFK, MTA Centre of Excellence, Konkoly-Thege Mikl\'os \'ut 15-17, 1121 Budapest, Hungary\\
$^{7}$ELTE E\"otv\"os Lor\'and University, Institute of Physics, P\'azm\'any P\'eter s\'et\'any 1/A, 1117 Budapest, Hungary\\
$^{8}${Department of Chemistry, Ludwig Maximilian University, Butenandtstr. 5-13, D-81377 Munich, Germany}
}

\date{Accepted 2023 November 28. Received 2023 November 28; in original form 2023 August 11}

\pubyear{2023}

\begin{document}
\label{firstpage}
\pagerange{\pageref{firstpage}--\pageref{lastpage}}
\maketitle

\begin{abstract}
FU~Ori-type stars are young stellar objects (YSOs) experiencing luminosity outbursts by a few orders of magnitude, which last for $\sim$$10^2$~years. A dozen of FUors are known up to date, but many more currently quiescent YSOs could have experienced such outbursts in the last  $\sim$$10^3$~years. To find observational signatures of possible past outbursts, we utilise ANDES, RADMC-3D code as well as CASA ALMA simulator to model the impact of the outburst on the physical and chemical structure of typical FU~Ori systems and how it translates to the molecular lines' fluxes. We identify several combinations of molecular lines that may trace past FU~Ori objects both with and without envelopes. The most promising outburst tracers from an observational perspective are the molecular flux combinations of the N$_{2}$H$^{+}$~$J=3-2$, C$^{18}$O~$J = 2-1$, H$_2$CO~$(J_{\rm K_a, K_c}) = 4_{04}-3_{03}$, and HCN~$J = 3-2$ lines. We analyse the processes leading to molecular flux changes and show that they are linked with either thermal desorption or enhanced chemical reactions in the molecular layer.
Using observed CO, HCN{, N$_2$H$^+$} and H$_2$CO line fluxes from the literature, we identify {ten} nearby disc systems that might have undergone FU~Ori outbursts in the past $\sim$$10^3$\,years: [MGM2012]~556, [MGM2012]~371 and [MGM2012]~907 YSOs in L1641, Class~II protoplanetary discs around CI~Tau, AS~209 and IM~Lup and transitional discs DM~Tau, {GM~Aur,} LkCa~15 and J1640-2130.
\end{abstract}

\begin{keywords}
protoplanetary discs -- circumstellar matter -- accretion -- astrochemistry -- stars: pre-main-sequence
\end{keywords}



\section{Introduction}
Some of the rare and notable objects at early stages of stellar evolution are the FU~Ori-type stars (FUors). These young stars are observed in a state of a luminosity outburst, which can reach hundreds of solar luminosities in amplitude and several decades or centuries in duration~\citep{2023ASPC..534..355F}. Meticulous analysis and modelling of the FUors' observations point at a temporal increase in the accretion rate as the cause of the outbursts~\citep{1985ApJ...299..462H}. Several potential mechanisms for the accretion variability have been proposed, including gravitational~{\citep{2011ApJ...729...42M, 2017MNRAS.464L..90M, 2015ApJ...805..115V}}, magneto-rotational~\citep{2009ApJ...701..620Z, 2020ApJ...895...41K}, and convective instabilities in the disc~\citep{2020ARep...64....1P}, or stellar flybys~\citep{2010MNRAS.402.1349F, 2022MNRAS.517.4436B}. However, the exact driving force behind these outbursts is not yet well understood.  

Understanding of the FU~Ori phenomenon is hampered by the low statistics, since only around ten confirmed FUors undergoing an outburst phase have been discovered to date~\citep{2014prpl.conf..387A, 2023ASPC..534..355F}.
{In addition, there are several possible FUors, namely V899~Mon, V346~Nor, V960 Mon and OO~Ser, which experienced a significant decrease or halt in their accretion~\citep{2023ASPC..534..355F}.}
However, there is still no consensus in the community of how an actual post-outburst FUor (post-FUor) system should look from the observational perspective.

A promising way for identifying the post-FUor systems is to search for the observational tracers which remain significantly altered with respect to the quiescent state for $\gtrsim$10--100~years. As previous theoretical studies have shown, the powerful outbursts affect disc chemical composition by evaporation of ices and thermal processing of the gas~\citep[e.g.,][]{2007JKAS...40...83L, 2012ApJ...754L..18V, 2012ApJ...758...38K, 2013A&A...557A..35V, 2017A&A...604A..15R, 2018ApJ...866...46M, 2019MNRAS.485.1843W}.
It has been found that the evaporated CO ice can remain in the gas phase in the envelope for several kyr, which is an order of magnitude longer than the typical outburst timescale~\citep{2012ApJ...754L..18V, 2013A&A...557A..35V}. The increased gas-phase abundance of CO leads to more efficient production of HCO$^+$ and other CO-related species, while suppressing the N$_2$H$^+$ and other protonated molecular ions' abundances on a comparable timescale~\citep[][]{2007JKAS...40...83L, 2012ApJ...754L..18V}. The absence of CO on dust can be likewise noticed in the dust emission~\citep{2012ApJ...758...38K}. 

In~\citet{2017A&A...604A..15R},  an approach for identification of embedded post-FUors was suggested based on the elevated C$^{18}$O emission. Non-CO molecular tracers were further explored in a paper by~\citet{2018ApJ...866...46M}, who used an extensive chemical network to study outburst chemistry in Class~II YSOs. It was found that in the post-FUor systems without the envelope, CO rapidly returns to the pre-outburst state, and H$_2$CO and NH$_2$OH were proposed as short-term and long-term tracers, respectively. \citet{2019MNRAS.485.1843W} studied the effects of the outburst on a variety of chemical processes in a number of single-point models representing different regions of the disc. They reproduced the conclusions regarding CO for the FUors with and without the envelope, and found that the outburst can trigger the effective formation and thermal desorption of complex organic molecules (COMs), such as HCOOCH$_3$, CH$_2$CO, CH$_3$CN, in the disc. Indeed, a number of COMs (CH$_3$OH, HCOOCH$_3$, CH$_3$CHO, HCOOH, CH$_3$COCH$_3$, HC$_3$N, etc.) were detected in the disc of an outbursting FU~Ori star V883~Ori~\citep{2019NatAs...3..314L, 2018ApJ...864L..23V}, in agreement with theoretical predictions of thermal desorption of complex ices and high-temperature gas-phase chemistry~\citep[e.g.][]{2016ApJ...821...46T,2018ApJ...866...46M}.  

In the present paper, we continue our previous theoretical studies of outburst chemistry~\citep{2018ApJ...866...46M, 2019MNRAS.485.1843W}, now with the aim at identifying promising observational tracers of the {\it past} FUor outbursts. As in \citet{2018ApJ...866...46M}, we utilise the physico-chemical disc model ANDES to simulate variable temperature structure and time-dependent chemistry~\citep{2013ApJ...766....8A}. It is coupled with the line radiative transfer code RADMC-3D and the ALMA simulator in CASA to derive synthetic line fluxes. 
In addition, we extend our research to both the Class~I sources with envelopes and the Class~II sources without them, while our previous work only focused on the latter. Observed FUors are represented by both of these classes~\citep{2007ApJ...668..359Q, 2017ApJ...836..226K}. Finally, we compare our theoretical results to the {molecular line} observations in a number of Class~I and II sources from the literature.

The paper is organised as follows. In Section~\ref{sec:meth}, we describe our methods, including our suggested post-FUor tracer definition and modelling setup. In Section~\ref{sec:res}, we present our results, namely, the proposed post-FUor identification criteria, chemo-physical explanations, and comparison with the observations. Finally, we discuss the applicability of the results to observations in Section~\ref{sec:dis} and summarise our work in Section~\ref{sec:concl}.

\section{Methods}\label{sec:meth}

\subsection{Disc physical structure}
To calculate the physical structure and chemical evolution of the circumstellar disc and the surrounding envelope, we use the ANDES physico-chemical code~\citep{2013ApJ...766....8A}. ANDES describes a 2D~axisymmetric protoplanetary disc in hydrostatic equilibrium in the cylindrical $(R,z)$ coordinates, where $R$ is the radial distance projection to the disc midplane and $z$ is the elevation above the midplane. We consider radial distances $R$ from 0.5 to 3\,000\,au, and aspect ratios $z/R$ from 0 to 2. In this work, we employ the version of the ANDES model adapted for treating luminosity outbursts~\citep{2018ApJ...866...46M}, with some further modifications. Below, we recap the main characteristics of the model, as well as newly added features.

Thermal and density structures in the vertical direction are defined self-consistently. Gas temperature is assumed to be equal to the dust temperature, which, in turn, is governed by the radiative heating from the central source. The temperature in the disc upper layers {$T_{\rm a}$} and the UV radiation field (originating in the hot accretion region) are calculated using 2D radiative transfer. {Midplane temperature $T_{\rm mp}$ is calculated in an optically thick approximation following Eq.~(8) in \citet{2017ApJ...849..130M}. It depends on the stellar luminosity defined by} the temperature and radius of the central star with a given mass at 1\,Myr age from the isochrones of~\citet{2008ASPC..387..189Y}, and on the accretion luminosity $L_{\rm acc}$, which is defined as an accretion region of a variable size with $T=10\,000$\,K in proximity of the star{, assuming blackbody radiation spectrum. The accretion luminosity is chosen so that it mimics the FUor outburst, its evolution is described in Section~\ref{sec:outburst_model}. It is also used to calculate the UV radiation field affecting the rates of photoreactions.} {The local temperature is determined parametrically from $T_{\rm a}$ and $T_{\rm mp}$ \citep[see Eq.~(5) in][]{2017ApJ...849..130M}.}

{The radial distribution of gas surface density is described by a tapered power law with an exponent $\gamma=1$~\citep[for more details see][]{2017ApJ...849..130M}. The parameterisation includes disc characteristic radius $R_{\rm c}$ and total gas mass in the disc $M_{\rm d}$. The vertical density structure is calculated from the assumption of hydrostatic equilibrium self-consistently with the temperature distribution.}

Dust grains comprise 1\% of the gas mass, and have a size distribution described by a power law with a~$-3.5$ exponent between $5\times10^{-7}$ and $2.5\times10^{-3}$\,cm. This corresponds to the mean dust size of $3.7\times10^{-5}$\,cm, slightly bigger compared to the pristine ISM dust grains. Here, mean dust size is the size of a grain if we replace our ensemble with monodisperse dust of the same total area and mass. {Settling of the grown dust to the midplane is not included in the model.} {The adopted dust distribution corresponds to ``medium'' dust from \citet{2018ApJ...866...46M} and represents moderately grown grains compared to the ISM. The choice of larger $a_{\rm max}$ would lead to further slowing down the surface chemistry, extending the timescales of the freeze-out and increasing the UV radiation field in the molecular layer. With these effects combined, \citet{2018ApJ...866...46M} found no clear effect of dust size on the chemical consequences of the FUor outburst. Although observations of protoplanetary discs indicate that dust grains grow to millimetre sizes in the disc midplane \citep{2014prpl.conf..339T}, we choose this moderately grown dust model to be able to represent dust in the disc upper layers, and the embedding envelope, as well as the midplane where small dust is also resupplied by collisional fragmentation~\citep{2012A&A...539A.148B}.} {Small grains provide the largest grain surface area (in case of the $-3.5$ exponent), which is the most important dust parameter for the surface chemistry modelling.}

Dust opacities are based on~\citet{1984ApJ...285...89D}. The calculation of the X-ray radiation field is adopted from~\citet{2009ApJS..183..179B}, with the assumed stellar X-ray luminosity $L_{\rm X}=10^{30}$\,erg~s$^{-1}$. The ionisation rate from the interstellar cosmic rays is calculated according to~\cite{2018A&A...614A.111P}, with the unattenuated rate $\zeta=10^{-17}$\,s$^{-1}$.

\subsection{Envelope modelling}
We added a surrounding envelope to our model of a protoplanetary disc to account for the embedded eruptive Class~I FUors. 
The implemented envelope model is based on the model by~\citet{2003ApJ...591.1049W} and ~\citet{2017A&A...604A..15R}. Gas density in the envelope at a distance $r$ from the star ($r^2=R^2+z^2$) is determined by the following expressions:
\begin{equation}
\rho_{\rm env}=\frac{\dot{M}_{*}\psi(\mu,r)}{4\pi r^{3/2}\left(2GM_{\star}\right)^{1/2}},
\label{eq:rhoenv}
\end{equation}
\begin{equation}
\psi(\mu,r)=\frac{\left(2/(1+\mu/\mu_{0})\right)^{1/2}}{\mu/\mu_{0}+2\mu_{0}^{2}R_{\rm c}/r}.
\label{eq:muenv}
\end{equation}
Here $\dot{M}_{*}$ is {the mass infall rate}, $R_{\rm c}$ is the centrifugal radius, which we assume equal to the disc characteristic radius which defines the exponential tapering  (see \citet{2017MNRAS.464L..90M} for details), $\mu=\cos{\theta}$ is the cosine of the angle at which the matter falls to the envelope relative to the disc axis, and $\mu_{0}$ is the value of $\mu$ at the infinity. The values of $\mu$ and $\mu_{0}$ are linked with each other as:
\begin{equation}
\mu_{0}^{3}+\mu_{0}(r/R_{\rm c}-1)-\mu(r/R_{\rm c})=0.
\label{eq:muenv1}
\end{equation}
{We adopt the mass infall rate of $5\times10^{-6}$\,$M_{\odot}$\,yr$^{-1}$ and the envelope radius of 3\,000\,au. We choose the value of $\dot{M}_{*}$ to obtain reasonable envelope masses, and this value is different from the mass accretion rate from the disc to the star $\dot{M}_{\rm acc}$, which produces the accretion luminosity $L_{\rm acc}$ (see Section~\ref{sec:outburst_model} for details). We combine the density distributions of the disc and the envelope adding their densities at every point. Then the temperature is calculated for this combined density distribution.}

The adopted parameters result in the envelope mass of about $0.078 \times (M_{\star}/M_{\sun})^{-0.5}$. For low-mass stars (a fraction of $M_{\odot}$), the chosen model produces massive embedding envelopes with masses comparable to the stellar mass. The resulting envelope masses around more massive stars ($M_{\star} > $ 1--2\,$M_{\odot}$) constitute only several percents of the mass of the central star.
In models without an envelope,
we use a constant floor density of $10^{-24}$\,g~cm$^{-3}$ for the surrounding tenuous gas.

\subsection{Disc chemical structure}
The chemical evolution is described by time-dependent chemical kinetics equations, using the ALCHEMIC reaction network~\citep{2011ApJS..196...25S} updated with the newer data from the KIDA database\footnote{\url{http://kida.obs.u-bordeaux1.fr/},~\citet{2015ApJS..217...20W}}. {The surface chemistry treatment was also updated compared with \citet{2018ApJ...866...46M},  but the same reaction network is used, except for the binding energies of some species updated according to the values from~\citet{2017SSRv..212....1C}.} The network involves 650~chemical species and 7807~reactions, including gas-phase and surface two-body reactions, adsorption and thermal/reactive desorption, photoreactions, and ionisation or dissociation by X-rays, cosmic rays, and radioactive nuclides. The adopted reactive desorption yield is~1\%.

{The changes in binding energies $E_{\rm B}$ concern a few species with typically high abundances in the ISM and in protoplanetary discs. As the ice mantles in protoplanetary discs mostly consist of water, we favoured the laboratory values obtained for water ice substrate or for pure species~\citep{2017SSRv..212....1C}. Binding energy for water remains unchanged at 5700\,K. Minor changes were implemented to main carbon bearing species: from 1150\,K to 1180\,K for CO, from 2580\,K to 2860\,K for CO$_2$.
The changes of $E_{\rm B}$ for H$_2$CO from 2050\,K to 3260\,K and for CH$_3$OH from 5530\,K to 4235\,K will have the effect on the abundances of these species, as well as those of more complex organic molecules  that are formed from them. Major nitrogen-bearing species are also affected with $E_{\rm B}$ updated from 5530\,K to 3075\,K for NH$_3$ and from 1000\,K to 800\,K for N$_2$. The most significant changes are made for elemental and molecular oxygen: from 800\,K to 1660\,K for O, from 1000\,K to 898\,K for O$_2$, which would affect all surface chemistry.}

We modified the chemical model to account for the thickness of icy mantles covering dust grains in order to limit the efficiency of certain surface processes to a few (upper) monolayers of ice. 
This concerns the two-body surface reactions and photo-desorption both by UV photons and cosmic ray induced UV photons, as these reactions do not proceed efficiently in the bulk ice mantles. To make our model more feasible in that regard, we restrict the rates of the aforementioned reactions by multiplying it by the reduction factor $q_{\rm red} = 1 - e^{-3/k_{\rm ml}}$. Here, $k_{\rm ml}$ is the number of monolayers of ice on a grain at a certain location and a time moment. Assuming that ice mantles are thin, we calculate $k_{\rm ml} = \dfrac{n_{\rm ice}}{ 4\pi {a_{\rm dust}}^2 N_{\rm ss}}$, using the total number of molecules on the surface of a dust grain $n_{\rm ice}$, the average dust size $a_{\rm dust}=3.7\times10^{-5}$\,cm, and the surface site density $N_{\rm ss}=4\times10^{14}$\,cm$^{-2}$.

{The observed low CO abundances point out that some depletion processes are happening in protoplanetary discs \citep{2016ApJ...828...46A, 2017MNRAS.464L..90M}. Apart from freeze-out and photo-dissociation, there is chemical depletion and sequestration with dust grains \citep{2020ApJ...899..134K}. Freeze-out and photo-dissociation are accounted for in our model, as well as the chemical depletion of CO from the gas due to its reprocessing into CO$_2$ ice in dust  grains \citep{2017ApJ...849..130M,2018A&A...618A.182B}. However, we do not consider dust settling and drift capable of moving CO to the disc midplane and to the star \citep{2017A&A...600A.140S,2018ApJ...864...78K}, or any additional parametric depletion of carbon \citep{2016A&A...592A..83K}. }

\subsection{Modelling of an outburst}
\label{sec:outburst_model}
\begin{figure}
    \includegraphics[width=\columnwidth]{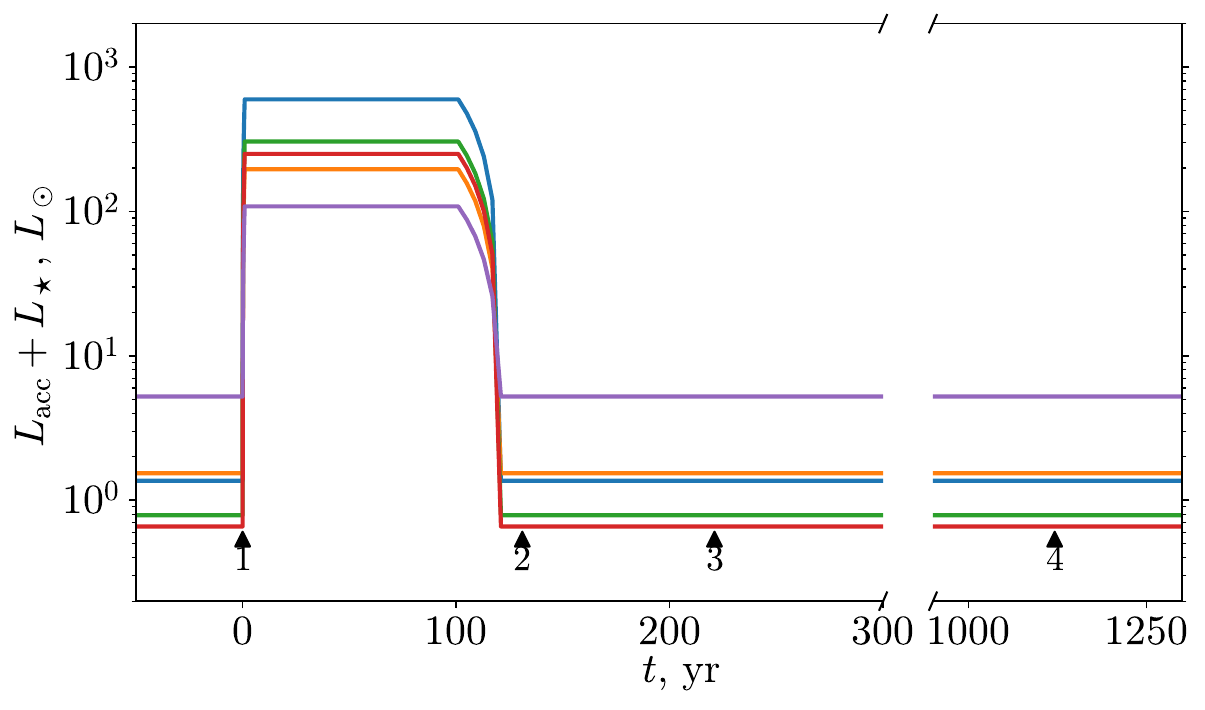}
    \caption{The evolution of the total luminosity of the central source (the sum of the stellar and accretion luminosity) for several models with different peak luminosity. Colours represent different models. Black triangles denote the key time moments: ($t_1$) the end of the quiescent pre-outburst phase ($t = 0$\,yr), ($t_2$)~10\,yr after the end of the outburst ($t = 131$\,yr), ($t_3$)~100\,yr after the end of the outburst ($t = 221$\,yr), and ($t_4$)~1000\,yr after the end of the outburst ($t = 1121$\,yr). }
    \label{fig:outburstprofile}
\end{figure}

To simulate the outburst, the accretion luminosity is varied according to the description of \cite{2018ApJ...866...46M}. We assume that the outburst begins after 500\,kyr in the quiescent state with $L_{\rm acc}=0.3$\,$L_{\sun}$, and that within one year $L_{\rm acc}$ grows linearly to the maximum value $L_{\rm burst}$. Then it remains in this high luminosity state for the next 100~years before linearly fading back to 0.3\,$L_{\sun}$ within 20~years. Thus, the total outburst duration is considered to be 121\,yr. Accretion luminosity evolution is set up manually to mimic the FUor luminosity profile. The chosen values of quiescent and peak luminosity correspond to mass accretion rate of {$\sim10^{-8}-10^{-7}$\,$M_\odot$~yr$^{-1}$ and $\sim10^{-5}-10^{-4}$\,$M_\odot$~yr$^{-1}$, respectively, given that the used expression $\dot{M}_{\rm acc} = L_{\rm acc}\dfrac{R_\star}{GM_\star}$ depends on stellar mass and radius varying across the model ensemble}.
The luminosity evolution for several models with different peak luminosity is shown in Figure~\ref{fig:outburstprofile}.

We assume that during the outburst, changes in the luminosity immediately affect the underlying disc physical structure. At a specific moment, the thermal, density, and radiation field distributions in the disc are calculated assuming a hydrostatic equilibrium based on current luminosity of the central source. This approach inherently implies that the timescales of the heat transport in the disc and hydrodynamical relaxation are shorter than both the duration of the outburst and the characteristic time of the luminosity change. 

While this assumption seems feasible in the inner disc, outside the inner $\sim$20\,au the dynamic timescales could become longer than the outburst duration \citep[see, e.g., Figure 3 in][]{2014ARep...58..522V}.
We verify the feasibility of this approach by estimating the relevant timescales in Appendix~\ref{ap:timescales}. We find that it is a reasonable assumption for most of the disc, albeit there are cases when the relaxation timescale can exceed the outburst duration. In this situation, the usage of a chemo-dynamical model would be more appropriate, especially for discs with high surface densities. Combining a hydrodynamic disc model with detailed gas-grain chemistry is a challenging task, which we refrain from doing in this study. 

During the outburst, the stellar X-ray luminosity $L_{\rm X}$ is assumed to be constant. It is yet unknown if FUor outbursts are accompanied by rising X-ray luminosity. There are several X-ray detections of FUors~\citep{2010ApJ...722.1654S,2014A&A...570L..11L}, and the difference in X-ray luminosity between FUors and regular YSOs is statistically significant~\citep{2019ApJ...883..117K}. But it remains unclear if the X-ray flux rises due to the outburst or is generally higher in the objects prone to the bursts~\citep{2019ApJ...883..117K}. The local FUV radiation field and ionisation rates in all disc cells are calculated by using the updated physical structure. Redistribution of gas and dust in the disc driven by the changing accretion luminosity is accounted for in the concentrations of all chemical species. For example, when the luminosity increases, the disc becomes more strongly heated, its local scale height increases, and the average gas density in the vertical direction decreases. In the envelope, dynamical timescales are long, so the density distribution does not change during the outburst. The envelope remains optically thin and acts as a bolometer, immediately heating up in response to the luminosity change.

The vertical distributions of abundances of all chemical species are updated accordingly at all radii. To calculate the new distribution of chemical species, we use the notion of a cumulative mass, which for a given radius is the species' mass below a given vertical height $z$. We assume that the cumulative mass of a species relative to the cumulative mass of the gas is preserved for all time moments. Thus, knowing the previous distributions of gas and the chemical species allows us to calculate the updated molecular concentrations from the updated gas density distribution. This approach allows accounting for the gas and dust advection necessary to reach the new quasi-equilibrium state when the luminosity of the central source varies. This abundance adjustment conserves the total amount of chemical species in the disc. Our disc model does not include other dynamical effects, such as turbulent mixing and advection.

{The adopted treatment of the vertical distribution of species is different from the one in \citet{2018ApJ...866...46M}, where the changes in physical structure were not accompanied by the corresponding adjustment of the chemical abundances. In the previous approach, the molecular layer remained at the same aspect ratio close to the midplane even when the vertical scale height increased during the outburst. The modification will allow to more carefully consider the evolution of the molecular layer under increased radiation field during the outburst. }

\subsection{Line radiative transfer modelling}

The physical and chemical disc structure from the ANDES model are then supplied to the RADMC-3D code~\citep{2012ascl.soft02015D} using the DiskCheF\footnote{\url{https://gitlab.com/SmirnGreg/diskchef}} package to obtain synthetic spectral cubes for the chosen molecular lines. We use the LTE assumption and the same spectral window of 20\,km~s$^{-1}$ with a 0.2\,km~s$^{-1}$ resolution in the line radiative transfer simulations. Disc component is assumed to rotate with Keplerian velocities and the envelope component is in free-fall (rotation of the envelope is neglected). The molecular line data are taken from the LAMDA database~\citep{2005A&A...432..369S}. The resulting ideal synthetic spectra are processed with the CASA ALMA {\tt simalma} routine to create synthetic ALMA datacubes~\citep{2022PASP..134k4501B}. The following {\tt simalma} parameters are used: the ALMA configuration is {\tt alma.cycle8.2} (second most compact configuration in Cycle 8), total integration time is $30$~min, {\tt pwv} is $0.47$, {\tt niter} $=50$, and {\tt mapsize} equal to the full image size. {Resulting beam size is around $1''\times1''$ for all lines}. The image centre and the disc centre are re-aligned by using coordinate corrections that are put in {\tt ptgfile}. The median of the spectrum outside the line emission is subtracted from the data in each pixel. Finally, the spectral cubes are integrated over the {whole emitting area} and the velocity range of the line emission to obtain the integrated line flux in units of mJy~km~s$^{-1}$. The resulting spectral noise level is $\sim$100\,mJy and it can accumulate an error in the flux up to around 100\,mJy~km~s$^{-1}$ for all lines.

These simulations are performed for several time moments indicated in Figure~\ref{fig:outburstprofile}, namely, right before the outburst, and then at 10, 100, and 1000\,yr after the end of the outburst. We do not consider later time moments because most chemical species would already return to their pre-outburst state \citep{2017A&A...604A..15R,2018ApJ...866...46M,2019MNRAS.485.1843W}. {Outburst time moments are not considered either because comparing them with observations is met with certain difficulties. Synthetic fluxes strongly depend on the thermal structure, which might be sufficiently dynamic during the outburst, while our model adopts quasi-stationary approach. The interpretation of the observations is also hampered by the presence of the outflows, which are not included/considered in our model \citep{2019ApJ...884..146T, 2023ApJ...945...80C}. Often integrated line fluxes are not even included in studies dedicated to FUor observations in molecular lines due to a complicated morphology and different lines tracing different parts of the object.}

\subsection{A grid of considered models}

To disentangle the effects of the outburst and the disc's physical structure on the resulting molecular abundances and line emission, we compute a grid of models by varying five key parameters: the stellar mass $M_\star$, the disc mass $M_{\rm d}$, the disc characteristic radius $R_{\rm c}$, the outburst luminosity $L_{\rm burst}$, and the disc inclination angle $i$. The adopted parameter ranges are specified in Table~\ref{tab:models}. The considered outburst luminosities are observationally motivated and have been compiled by \cite{2014prpl.conf..387A}. Typical disc and stellar masses, and the disc sizes for the low-mass YSOs are considered \cite{2011ARA&A..49...67W}. The lower limit of 20\,au for disc size is also motivated by recent observations of FUors, which typically have compact discs~\citep{2021ApJS..256...30K}. The higher limit of 0.1 for $M_{\rm d}/M_\star$ is chosen due to the disc model being quasi-stationary and thus not suitable to describe self-gravitating discs ($M_{\rm d}/M_\star > 0.1$). The inclination angles are varied between 0 and $90\degr$.

The resulting grid consists of 100 disc models equally split between discs with and without the envelope. Each disc model calculation is set up by randomly drawing the values of the five key parameters values from the specified ranges, assuming the uniform distribution.

\begin{table}
    \centering
    \caption{Ranges of selected model parameters}
    \begin{tabular}{cc}
\hline \hline
Parameter & Values\\\hline
$L_{\rm burst}$ & $100$ -- $600\, L_\odot$ \\
$M_\star$ & $0.1$ -- $2\,M_\odot$ \\
$M_{\rm d}/M_\star$ & $0.001$ -- $0.1$ \\
$R_{\rm c}$ & $20$ -- $200$\,au \\ 
$i$ & $0\text{\textdegree}$ -- $90\text{\textdegree}$\\ \hline
    \end{tabular}
    \label{tab:models}
\end{table}

\subsection{Molecular tracers of the past FU~Ori outbursts}\label{sec:tracersdef}

To identify which molecular lines may be sensitive to the past FUor outbursts, we focus on several millimetre transitions that are easily observed in protoplanetary discs, see Table~\ref{tab:lines}. There are other commonly observed transitions (e.g. CO $J = 1-0$, $3-2$ etc.), but we focus on a single representative transition per molecule due to demanding radiative transfer calculations. Many other molecular lines, such as the lines of COMs, can potentially be good tracers, however in this work we stop at the most common ones. For convenience of notation, we hereafter refer only to the molecule instead of the specific transition, as we model only one line per molecule.   

After obtaining the synthetic line fluxes, we compare them before and after the outburst for various line pairs. We consider a line pair as a tracer of the past FUor outbursts if the modelled states before and after the outburst separate into distinct groups on the flux-flux diagram. A region in the flux-flux plot where only the post-outburst values lie we call a post-FUor area. If disc observations with their error bars appear in the post-FUor area in the corresponding flux-flux plot, such disc is considered as a post-FUor candidate. We note that the criteria defining this post-FUor area are model-dependent, and under different assumptions and for different models other line pairs may become sensitive to the FUor outbursts. The most promising pairs for identification of the post-FUors would consist of one line with an elevated flux after the outburst, another line whose flux is not affected by or is decreased due to the outburst, and with both lines scaling in the same manner with the disc mass, size, or quiescent temperature. 

\begin{table}
	\centering
	\caption{A list of molecular transitions studied as a potential tracers of post-FUor activity. The second column gives examples of the disc studies where these lines have been observed (for any YSOs status).}
	\label{tab:lines}
	\begin{tabular}{cl} 
		\hline
		Molecular line & Sources \\
		\hline
		\multirow{2}{\widthof{CO $J=2-1$}}{CO $J = 2-1$} & {\cite{2021ApJS..257....5Z}, \cite{2020PhDT........11B}},\\
         & {\cite{2019arXiv190311868D}, \cite{2018ApJ...863..106K}}\\\hline 
		\multirow{2}{\widthof{$^{13}$CO $J = 2-1$}}{$^{13}$CO $J = 2-1$} & {\cite{2020PhDT........11B}, \cite{2019A&A...623A.150V}}\\
         & {\cite{2018ApJ...863..106K}, \cite{2021ApJS..257....5Z}}\\\hline
		\multirow{2}{\widthof{C$^{18}$O $J = 2-1$}}{C$^{18}$O $J = 2-1$} & {\cite{2020PhDT........11B}, \cite{2019arXiv190311868D}}\\
         & {\cite{2018ApJ...863..106K}, \cite{2021ApJS..257....5Z}}\\\hline
		\multirow{2}{\widthof{H$_2$CO $(J_{\rm K_a, K_c}) = 4_{04}-3_{03}$}}{H$_2$CO $(J_{\rm K_a, K_c}) = 4_{04}-3_{03}$} & {\cite{2020ApJ...893..101L}, \cite{2020ApJ...890..142P}}\\
         & {\cite{2018ApJ...863..106K}}\\\hline
		\multirow{2}{\widthof{HCO$^+$ $J = 4-3$}}{HCO$^+$ $J = 4-3$} & {\cite{2018AAS...23114709P}, \cite{2014ApJ...796..120W}}\\
        & {\cite{2020PhDT........11B}, \cite{2020ApJ...894...74B}}\\\hline
		\multirow{2}{\widthof{CS $J = 7-6$}}{CS $J = 7-6$} & {\cite{2018AAS...23114709P}, \cite{2021ApJ...914..113N}}\\
         & {\cite{2014ApJ...796..120W}}\\\hline
		\multirow{2}{\widthof{HCN $J = 3-2$}}{HCN $J = 3-2$} & {\cite{2021ApJS..257....6G}},\\
            & {\cite{2017ApJ...835..231H}}\\\hline
		\multirow{2}{\widthof{N$_2$H$^+$ $J = 3-2$}}{N$_2$H$^+$ $J = 3-2$} & {\cite{2013Sci...341..630Q,2013ApJ...765...34Q,2015ApJ...813..128Q,2019ApJ...882..160Q}},\\
            & {\cite{2007A&A...464..615D}}\\
		\hline
	\end{tabular}
\end{table}

\section{Results}\label{sec:res}

\subsection{Post-FUor line fluxes}\label{sec:pfflux}

For a randomly generated set of disc models, the integrated line fluxes were calculated for the specific time moments ($t_1$, $t_2$, $t_3$, $t_4$, see Figure~\ref{fig:outburstprofile}), assuming the same distance of 150~pc. Then, they were combined into all possible line pairs and plotted on flux-flux diagrams. {Out of 28 possible unique line pairs (choosing two lines from eight considered transitions, $C_8^2 = 28$) we find line pairs that can be called tracers: 10 in case of embedded discs and 21 in case of non-embedded discs. The order of lines in a pair was chosen so the horizontal axis flux changes little or decreases due to the outburst, while the vertical axis flux changes significantly (either rising or dropping).}
Figures~\ref{fig:no_env_tracers1}-\ref{fig:no_env_tracers2} and~\ref{fig:env_tracers} show flux-flux diagrams of the selected line pairs for non-embedded and embedded discs, respectively. {One can easily see there the separation of pre- and post-outburst values necessary for the proposed definition of a post-FUor tracer (Section~\ref{sec:tracersdef}).}
We can note general tendencies of the flux change due to the outburst for different lines. For all the considered lines, with the exception of CO and HCN, the outburst significantly alters (more than one order of magnitude) the line flux in at least one (embedded or disc-only) group of models. Any non-relaxational post-outburst flux change, if occurs at all, is mostly of small amplitude and always eventually succumbs to relaxation. This univocal evolution and absence of flux oscillations is what essentially allows us to draw the post-FUor areas.

\begin{figure*}
	\includegraphics[width=0.66\columnwidth]{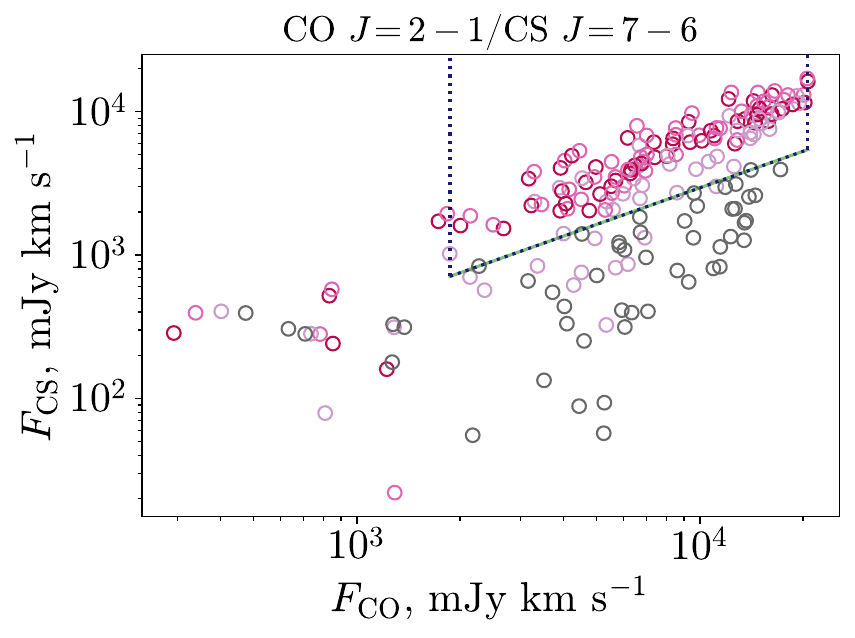}
	\includegraphics[width=0.66\columnwidth]{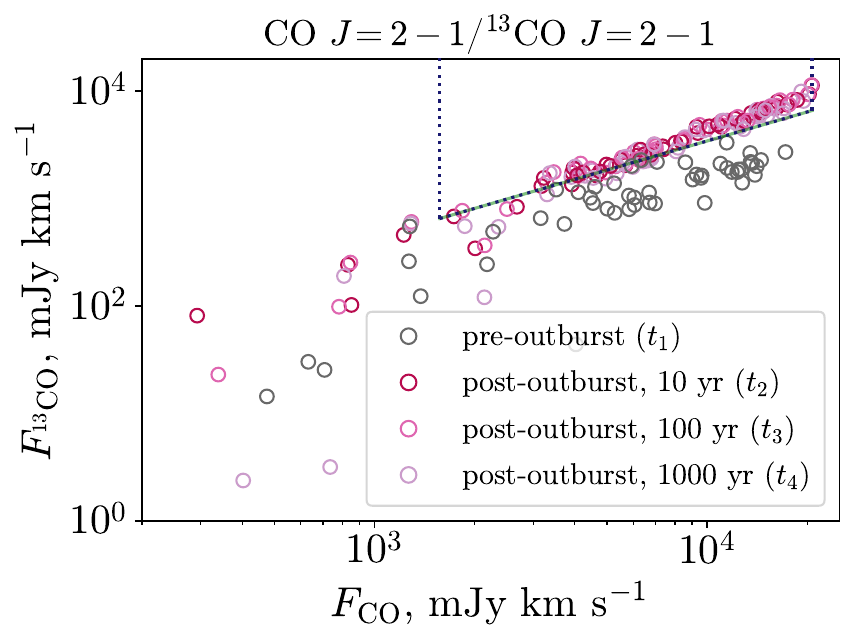}
	\includegraphics[width=0.66\columnwidth]{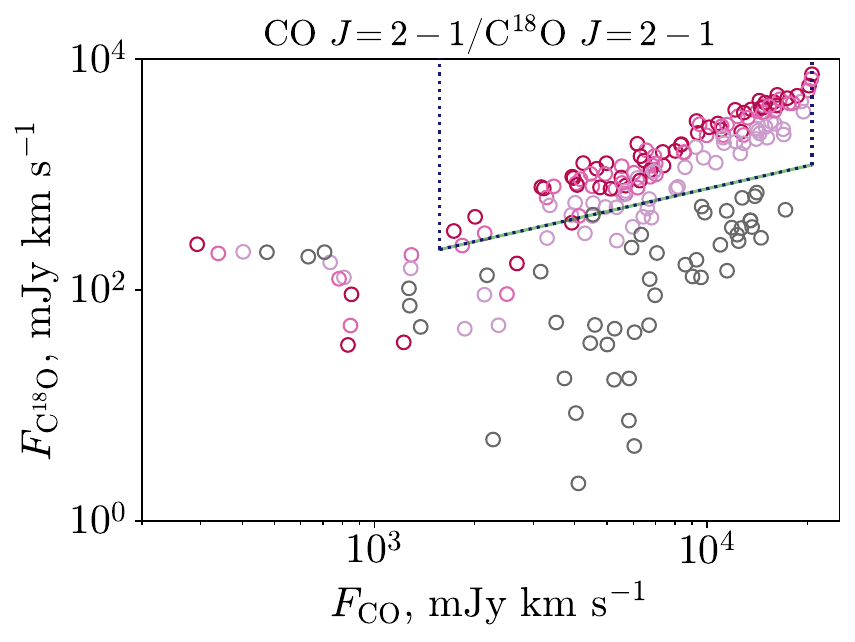}
	\includegraphics[width=0.66\columnwidth]{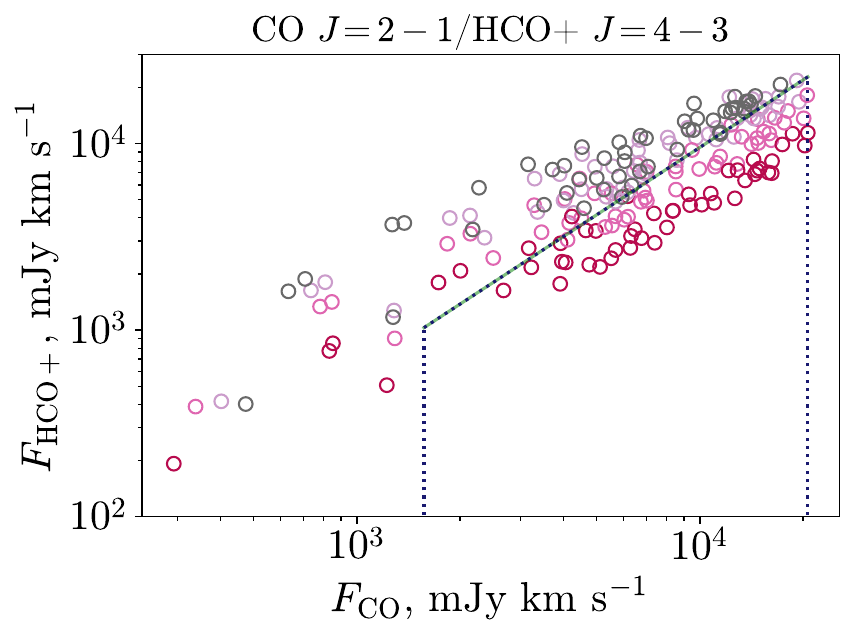}
	\includegraphics[width=0.66\columnwidth]{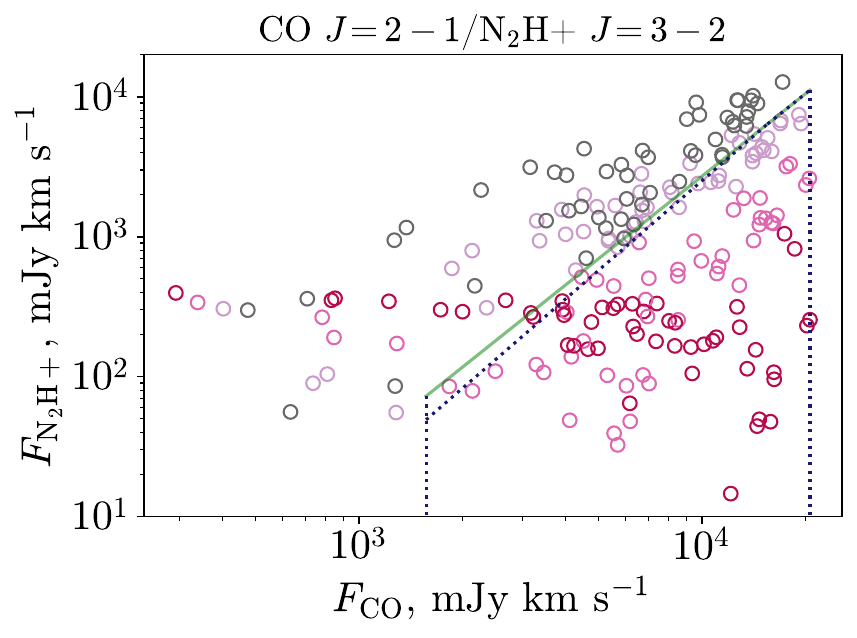}
	\includegraphics[width=0.66\columnwidth]{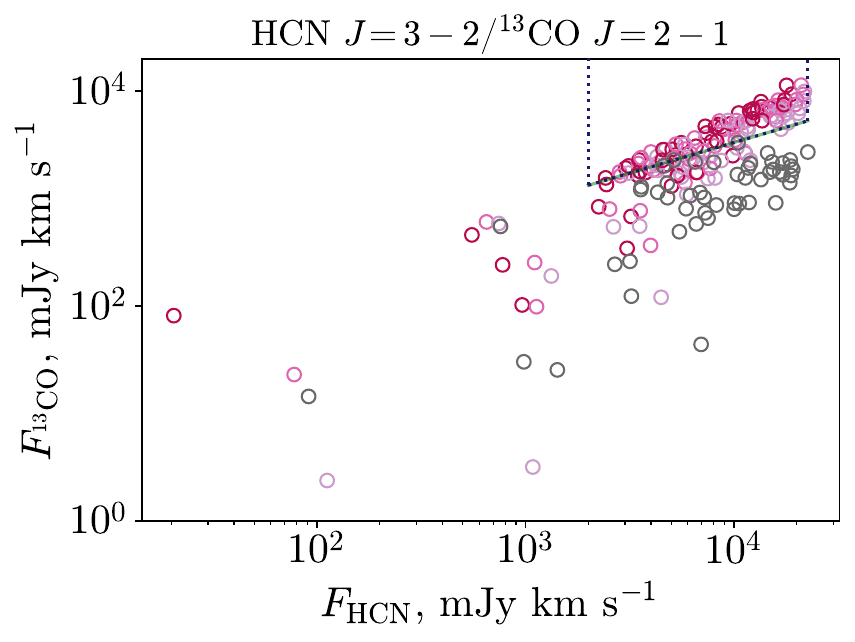}
	\includegraphics[width=0.66\columnwidth]{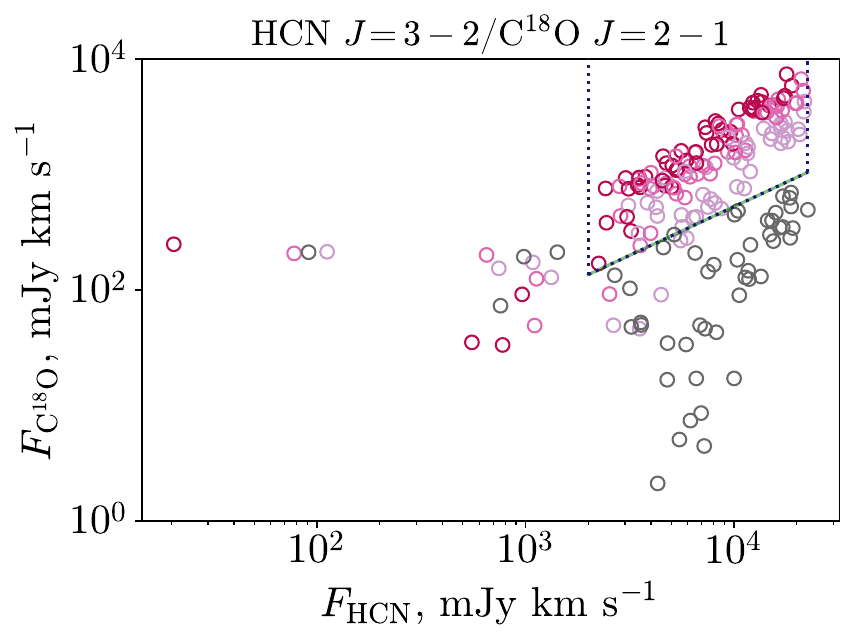}
	\includegraphics[width=0.66\columnwidth]{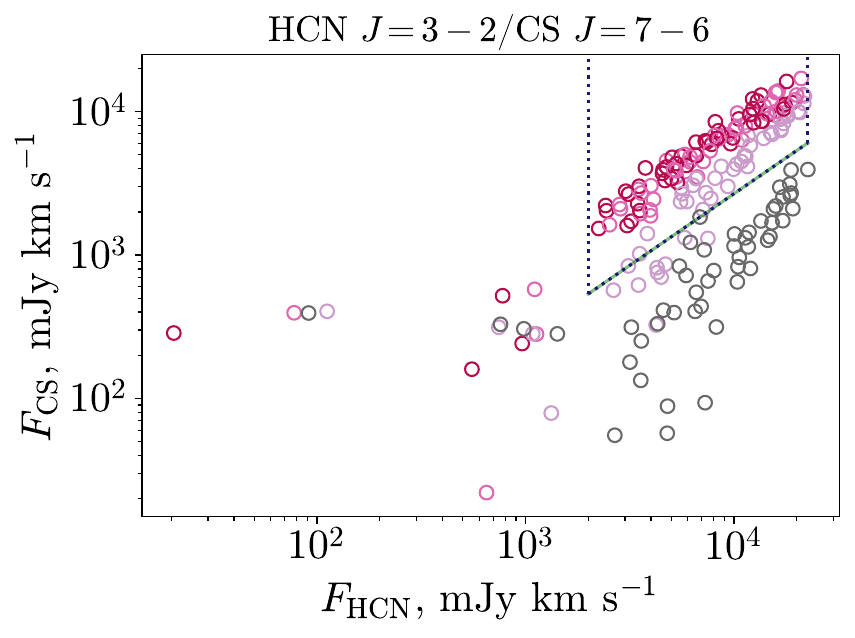}
	\includegraphics[width=0.66\columnwidth]{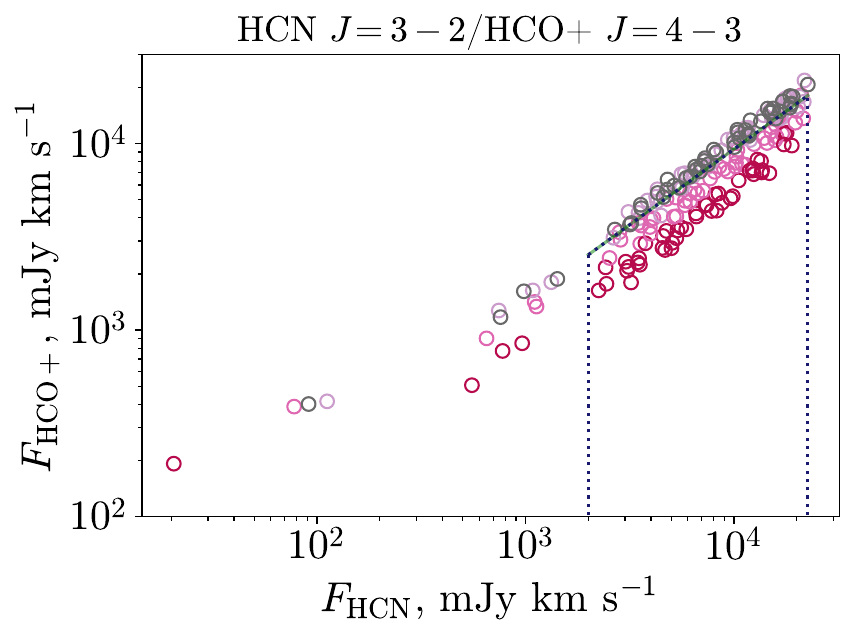}
	\includegraphics[width=0.66\columnwidth]{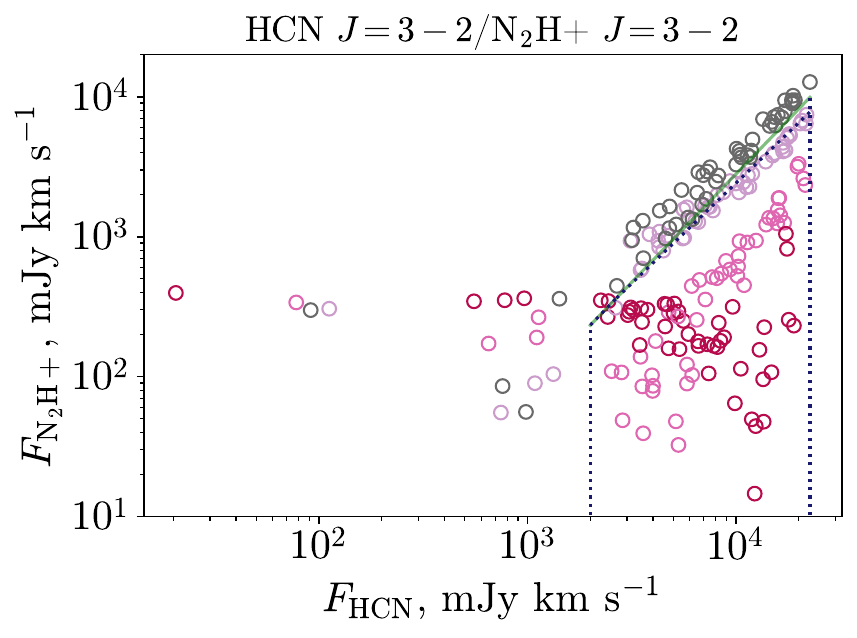}
	\includegraphics[width=0.66\columnwidth]{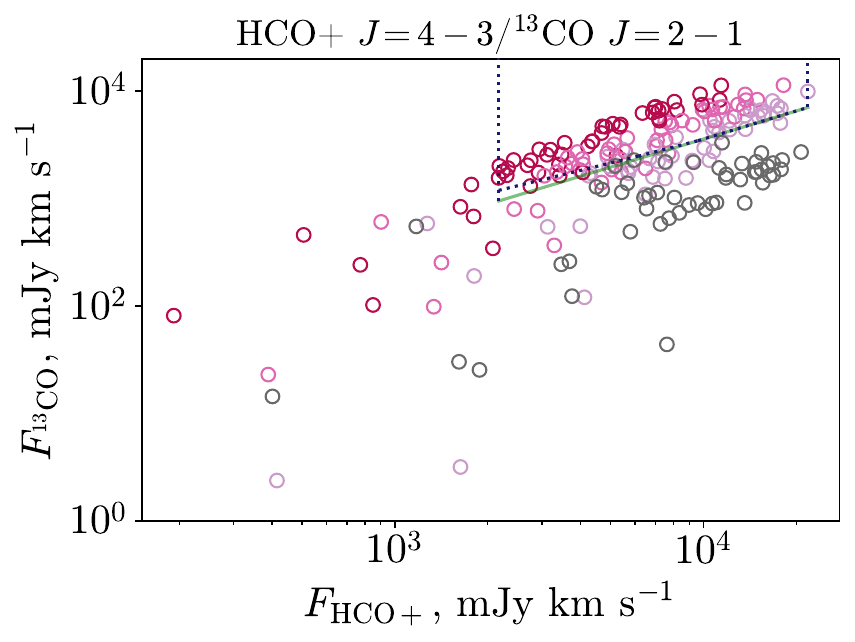}
	\includegraphics[width=0.66\columnwidth]{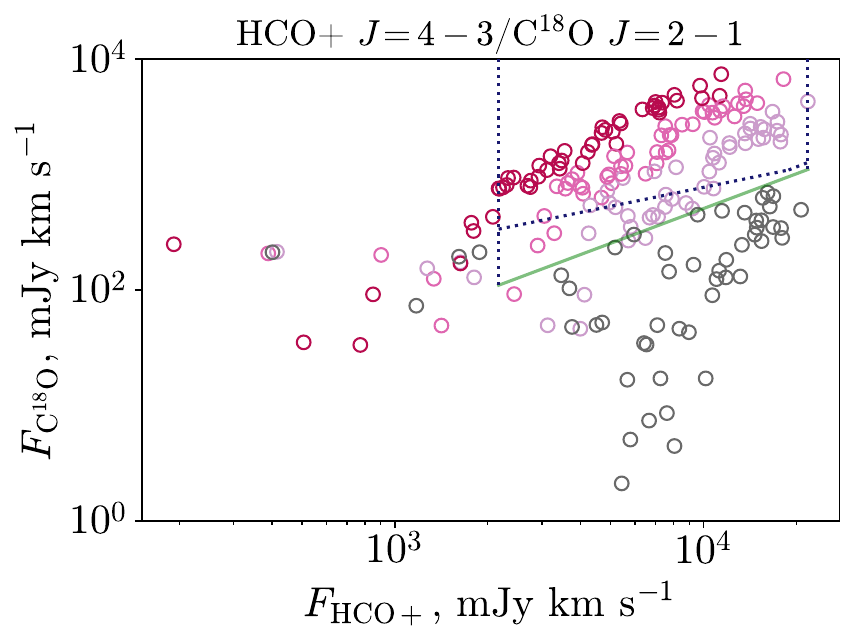}
	\includegraphics[width=0.66\columnwidth]{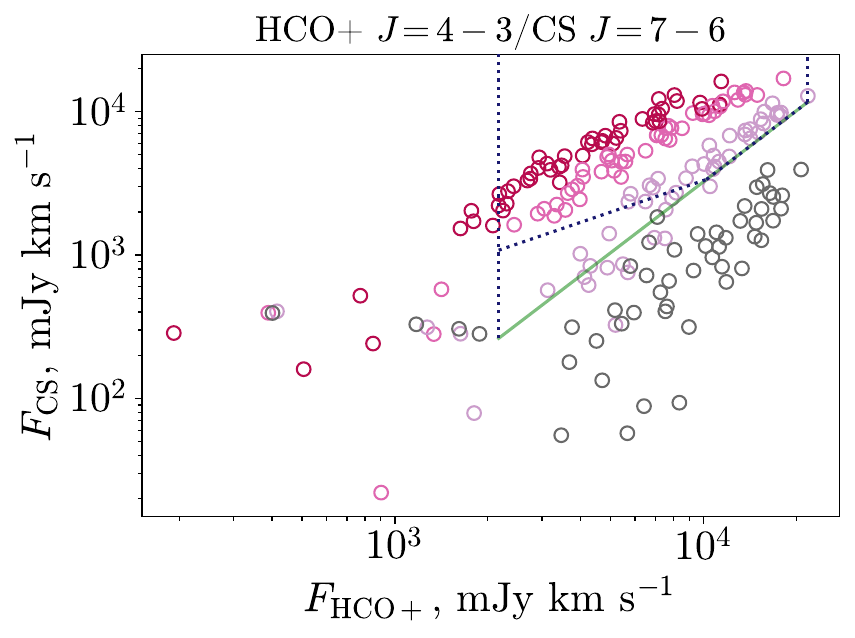}
	\includegraphics[width=0.66\columnwidth]{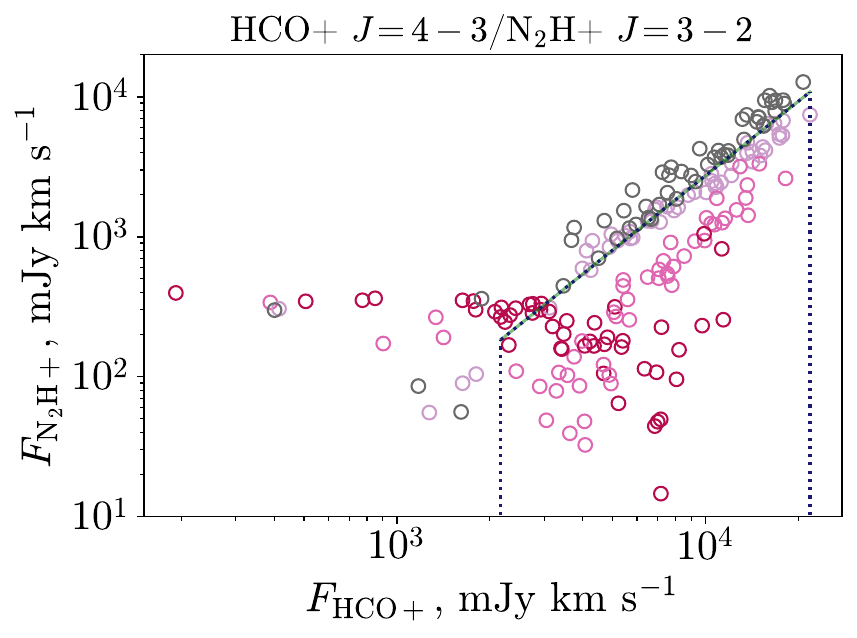}
	\includegraphics[width=0.66\columnwidth]{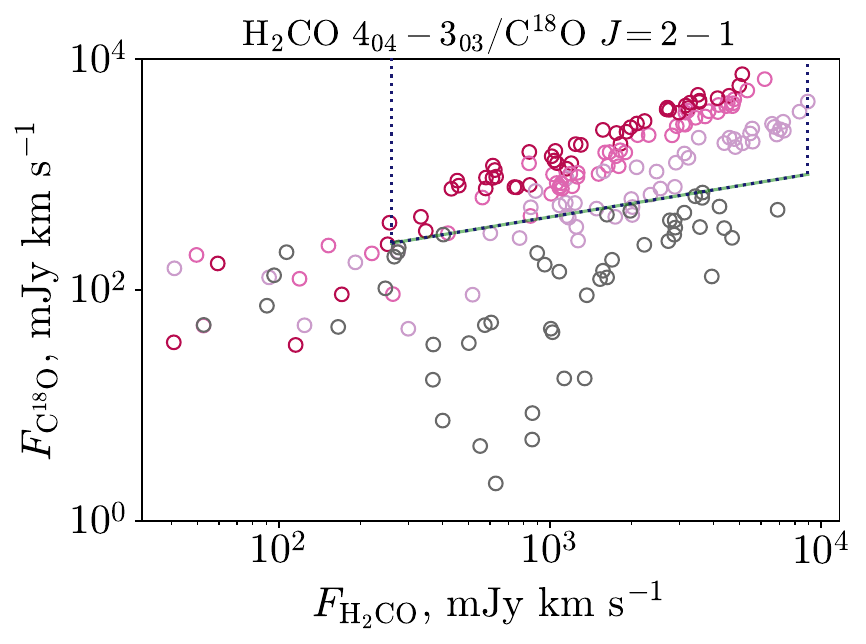}
    \caption{Flux-flux diagrams for several line pairs for the disc models with no envelope. Coloured markers designate the modelled discs at different evolutionary stages: grey for the pre-outburst ($t_1$) systems, dark red for the systems at 10\,yr after the outburst ($t_2$), pink for the systems at 100\,yr after the outburst ($t_3$), and lavender for the systems at 1000\,yr after the outburst ($t_4$). The time moments are indicated according to Figure~\ref{fig:outburstprofile}. Green translucent lines delineate pre-outburst and post-outburst flux values when only the corresponding pair of lines is considered. The dark blue dotted lines correspond to the combined criterion, which takes all these line pairs into account.}
    \label{fig:no_env_tracers1}
\end{figure*}

\begin{figure*}
	\includegraphics[width=0.66\columnwidth]{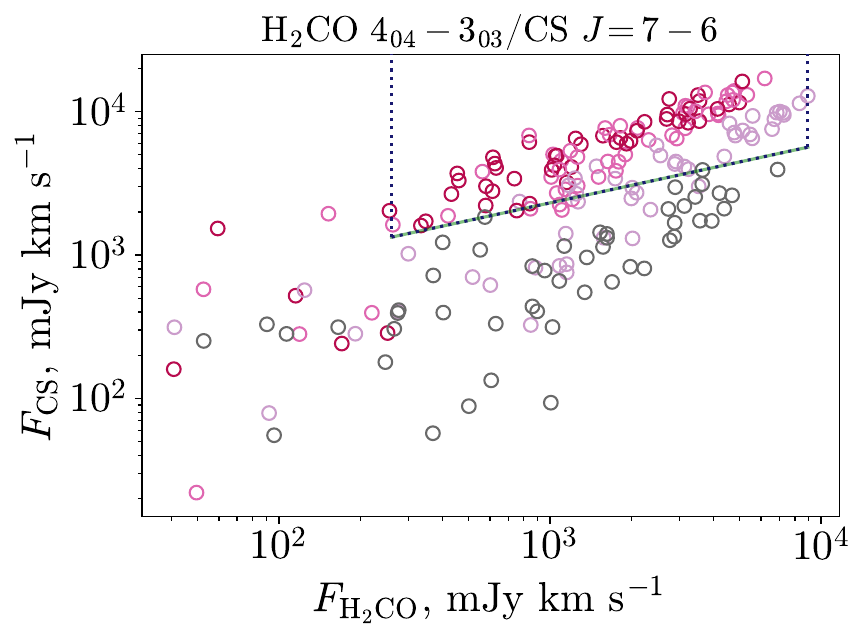}
	\includegraphics[width=0.66\columnwidth]{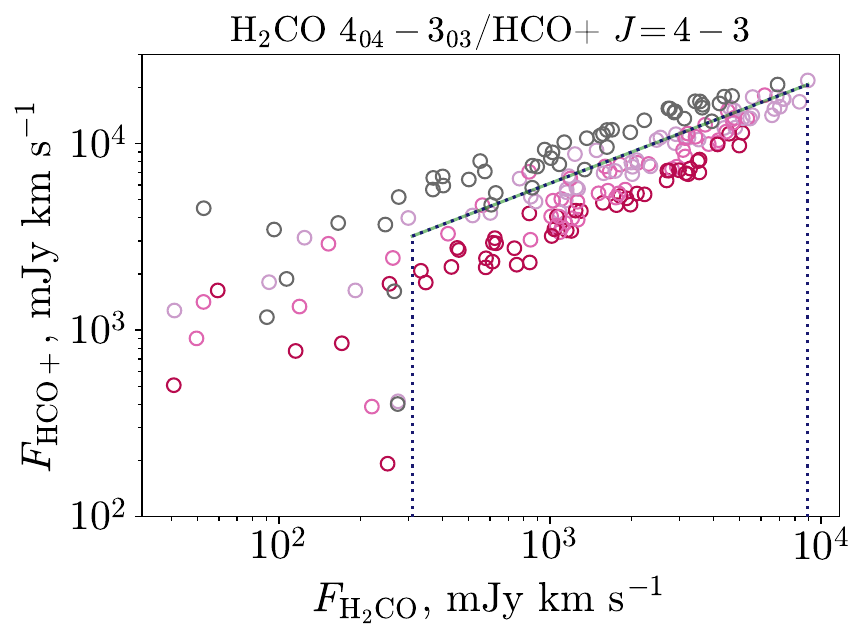}
	\includegraphics[width=0.66\columnwidth]{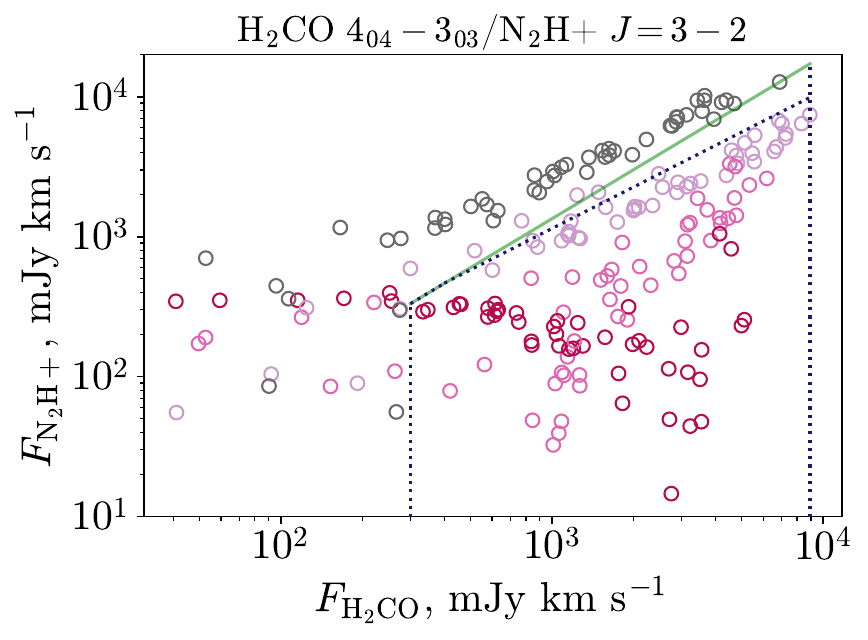}
	\includegraphics[width=0.66\columnwidth]{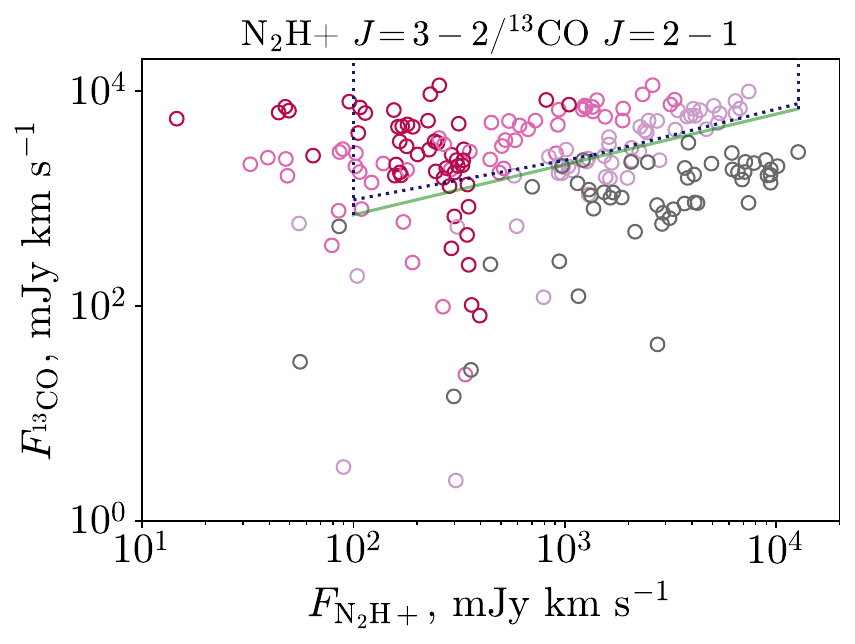}
	\includegraphics[width=0.66\columnwidth]{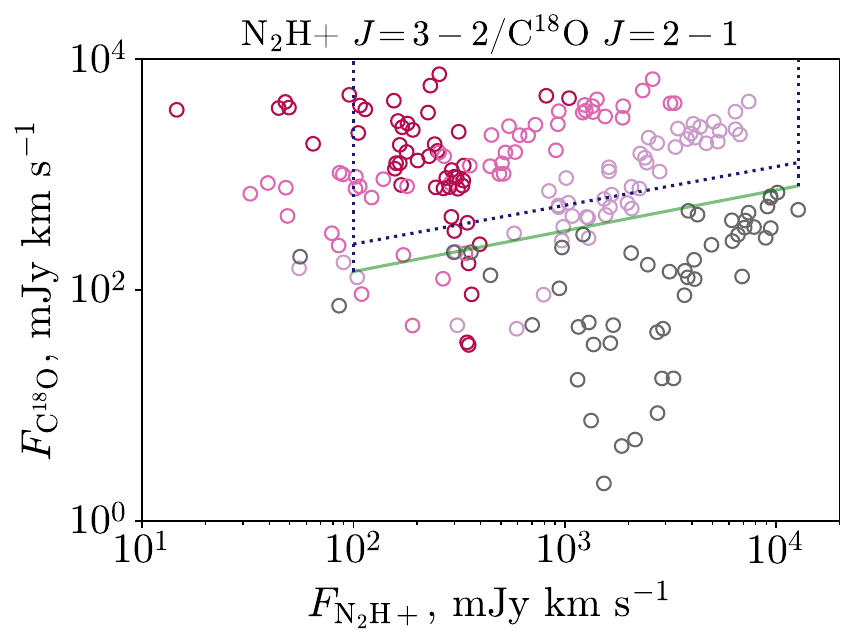}
	\includegraphics[width=0.66\columnwidth]{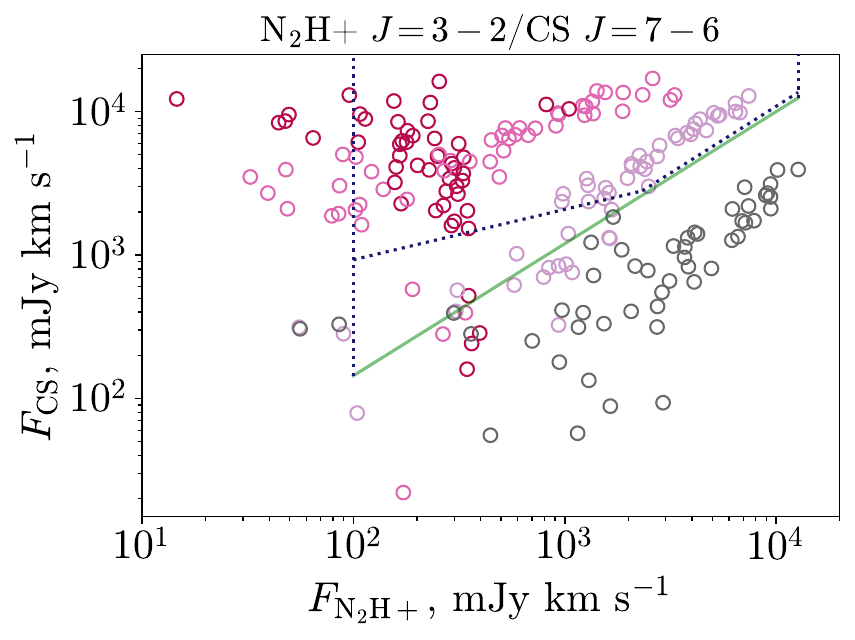}
    \caption{Continuation of Figure~\ref{fig:no_env_tracers1}.}
    \label{fig:no_env_tracers2}
\end{figure*}

\begin{figure*}

	\includegraphics[width=0.66\columnwidth]{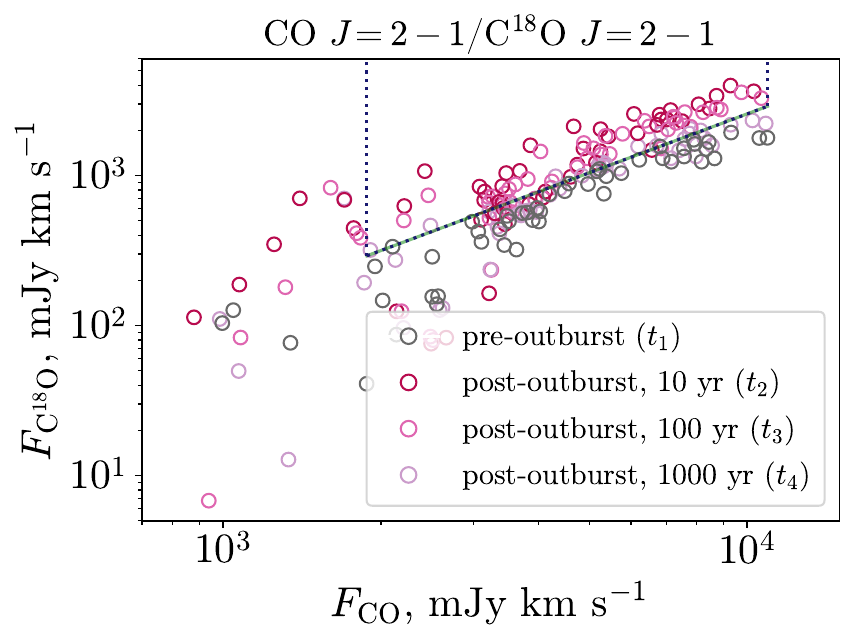}
	\includegraphics[width=0.66\columnwidth]{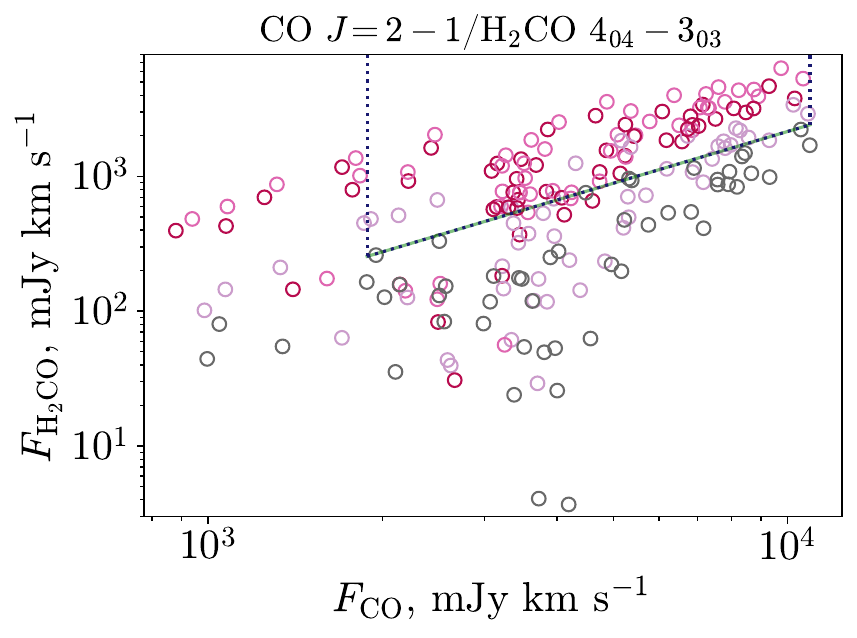}
	\includegraphics[width=0.66\columnwidth]{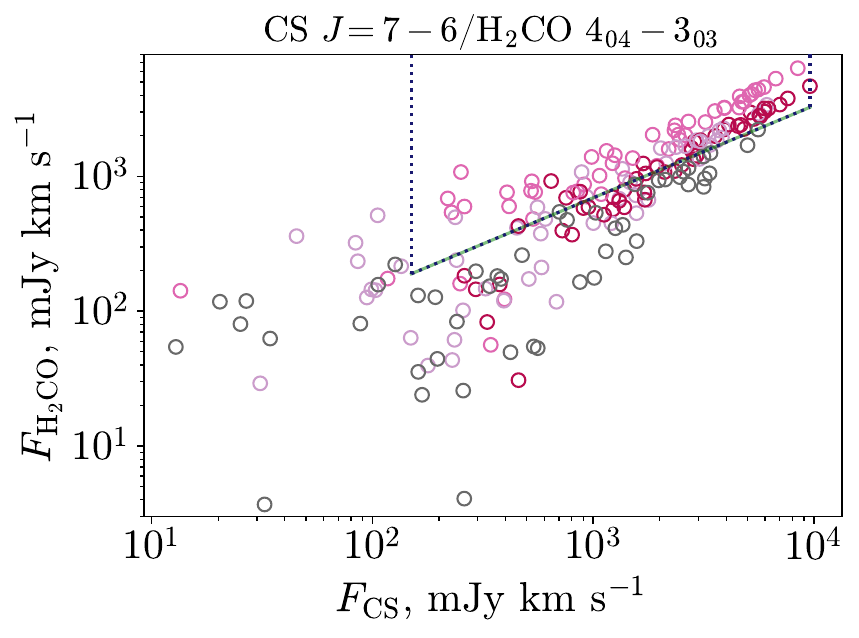}
	\includegraphics[width=0.66\columnwidth]{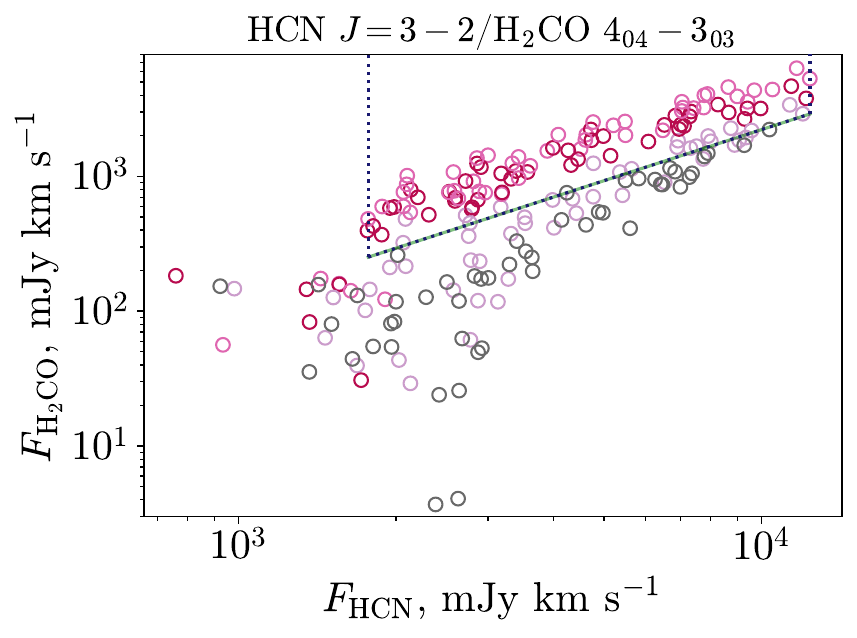}
	\includegraphics[width=0.66\columnwidth]{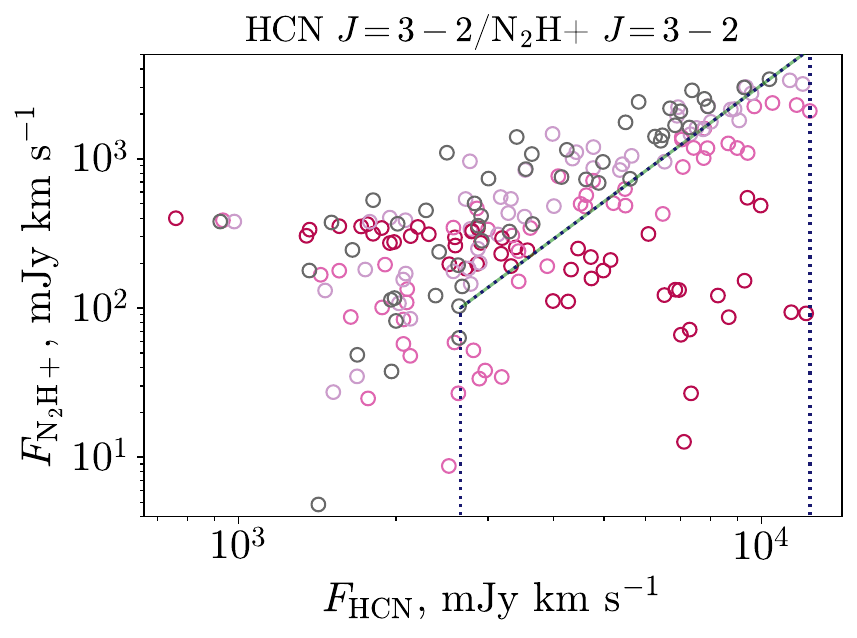}
	\includegraphics[width=0.66\columnwidth]{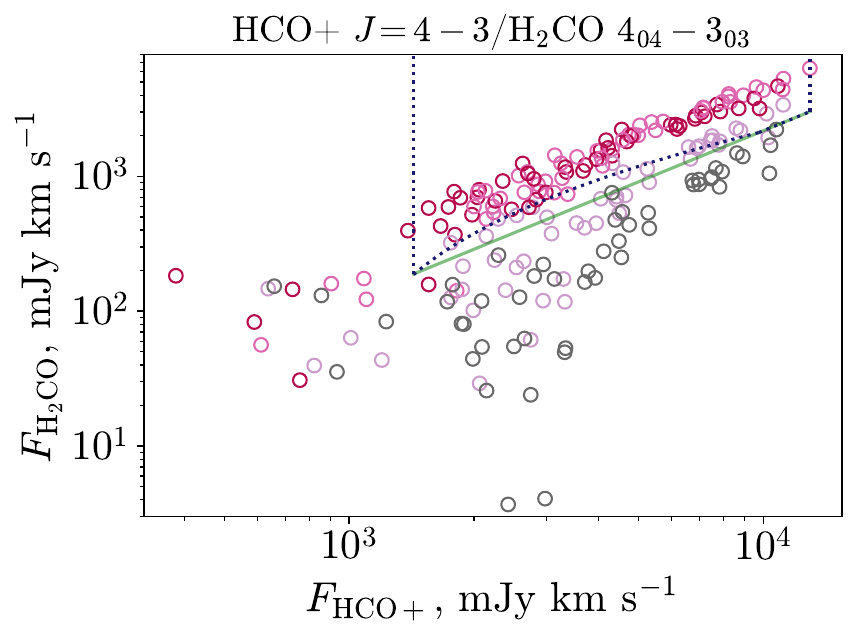}
	\includegraphics[width=0.66\columnwidth]{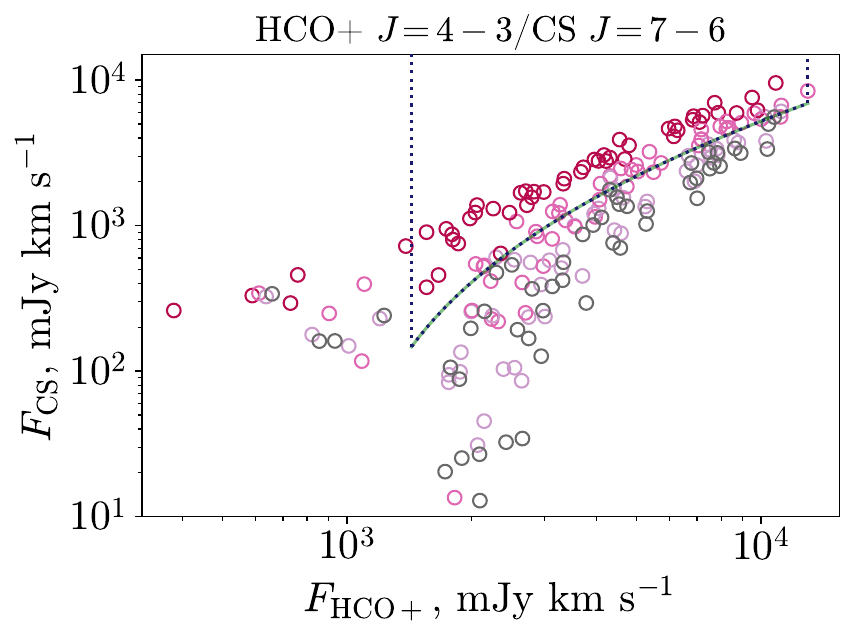}
	\includegraphics[width=0.66\columnwidth]{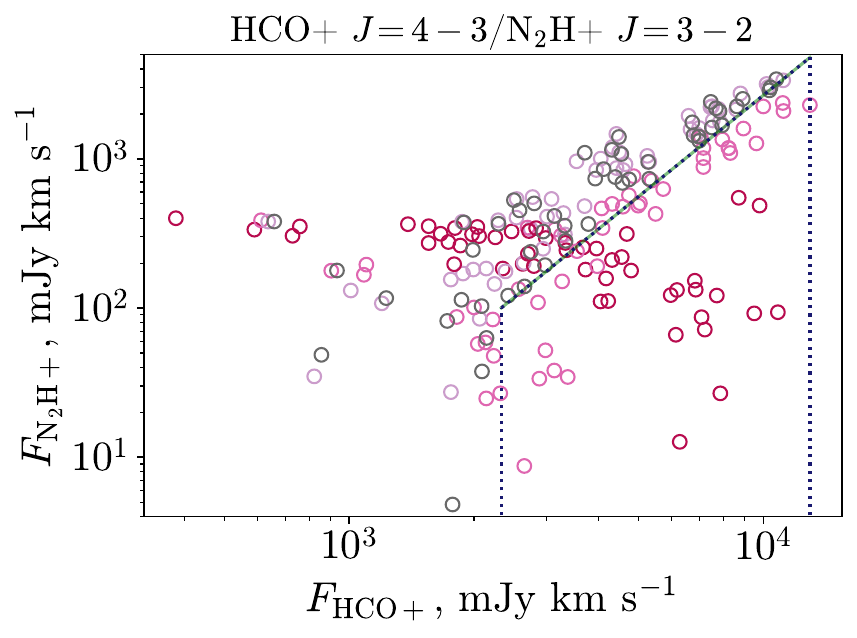}
	\includegraphics[width=0.66\columnwidth]{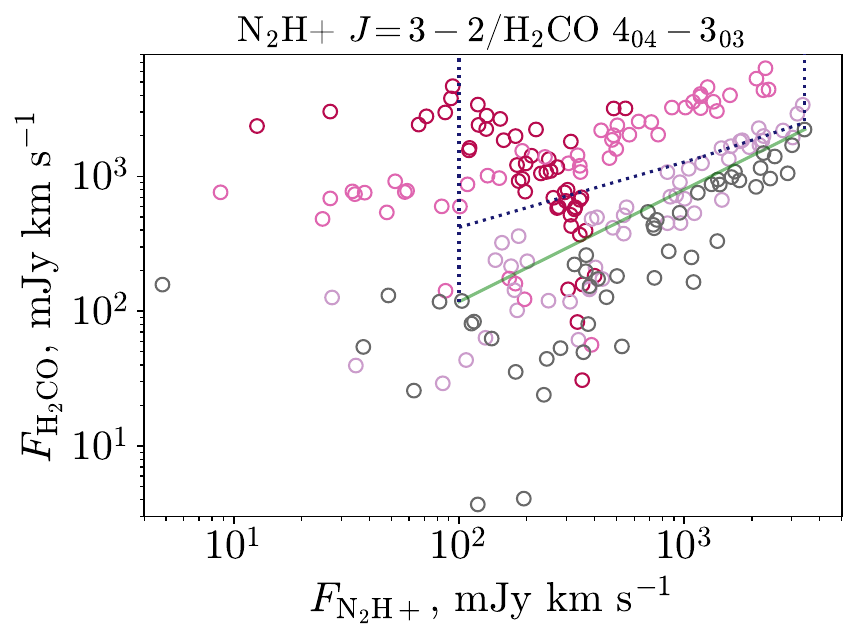}
	\includegraphics[width=0.66\columnwidth]{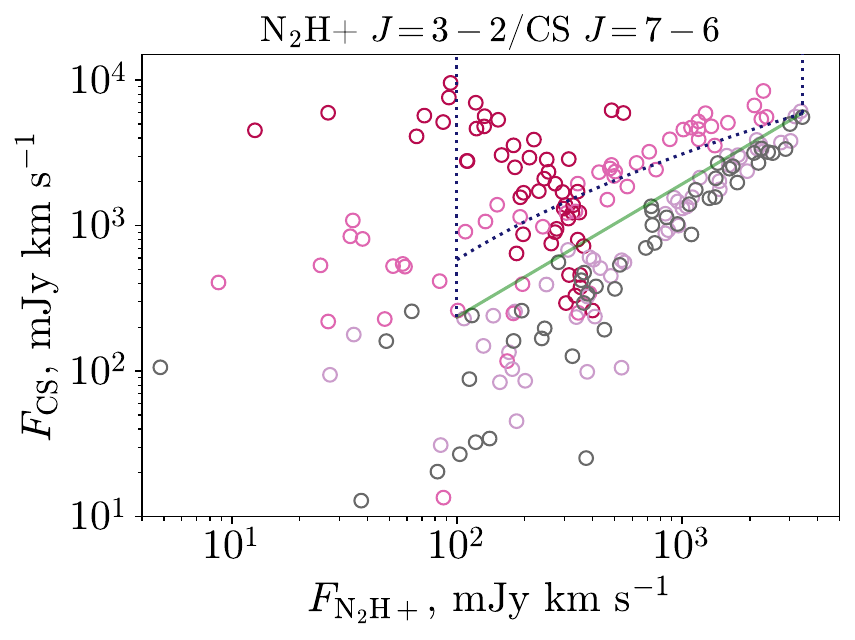}
    \caption{Same as Figure~\ref{fig:no_env_tracers1}, but for the models with an envelope.}
    \label{fig:env_tracers}
\end{figure*}

The particular behaviour of line fluxes depends on the presence of an envelope. First, we summarise flux evolution in the disc-only models (Figure~\ref{fig:no_env_tracers1} and \ref{fig:no_env_tracers2}). The line fluxes of $^{13}$CO and C$^{18}$O always increase between the pre-outburst state and the moment 10\,yr after it. After that they {need more than 1000\,yr to} fall back to the pre-outburst values. Line flux of CS shows similar rise, but their flux relaxation does not start until more than 100\,yr after the outburst. The H$_2$CO flux shows the behaviour similar to CS, yet at a much smaller amplitude. Lines of CO and HCN look mainly indifferent to the outburst. The HCO$^+$ {and N$_2$H$^+$} {fluxes decrease} during the outburst {and later return to the pre-outburst values in 1000\,yr}. Going to embedded models (Figure~\ref{fig:env_tracers}), we see mostly the same flux dynamics but at smaller magnitude. Exceptions are CS and H$_2$CO, which demonstrate opposite tendencies: H$_2$CO exhibits a significant flux rise while CS less so. Also the HCO$^+$ flux in this case appears indifferent to the outburst.

The flux evolution described above is true only for the majority of the discs with medium to high values of objects' masses and radii. For low-mass and compact discs, the molecular fluxes are lower and flux evolution is qualitatively different. Low-mass discs are more transparent in UV and their composition is stronger affected by photo chemistry. In compact discs, the snowlines of major volatiles might lay in the very outer diluted regions, so thermal desorption caused by increased accretion luminosity is irrelevant. Overall, the impact of outburst on the chemistry of low-mass compact discs is quite diverse depending on the specific molecule. On top of that, {in some lines (e.g. C$^{18}$O) these discs have too low flux to be detected, with signal-to-noise $\sim1$.} This does not allow us to draw any meaningful conclusions on post-outburst line flux evolution in low-mass compact discs.

\subsection{Past outburst criteria}
  
To describe the post-FUor area analytically, we use the relation between two integrated line fluxes of the following kind: 
\begin{equation}
\label{eq:ex}
    \log_{10}{f_j} = a_{ij} + b_{ij} (\log_{10}{f_i} - c_{ij})^{p_{ij}}
\end{equation}
where {$f_i = \dfrac{F_i}{1\text{ mJy km/s}}$, $f_j = \dfrac{F_j}{1\text{ mJy km/s}}$}, $F_i$ is the integrated $i$ line flux on the horizontal axis of the flux-flux diagram, $F_j$ is the integrated $j$ line flux on the vertical axis, and $a_{ij}$, $b_{ij}$, $c_{ij}$, and $p_{ij}$ are relation constants for a given line pair. For values of $c_{ij}$ and ${p_{ij}}$ the simplest case of linear relation ($c_{ij} = 0$ and ${p_{ij}} = 1$) is assumed by default. However, if the dependence on model parameters of pre- or post-outburst values noticeably deviates from linear, we tweak $c_{ij}$ and $p_{ij}$ to approximately follow the dependence. Values of $b_{ij}$ are derived by minimising the number of post-outburst values below the line described by relation (or above it if the flux is decreasing during the outburst like in the case of HCO$^+${and N$_2$H$^+$}). Minimisation is performed by counting post-outburst values that do not satisfy the criterion as a function of $b_{ij}$ and searching for the global minimum of this function using \texttt{scipy.optimize.basinhopping}\footnote{\url{https://docs.scipy.org/doc/scipy/reference/generated/scipy.optimize.basinhopping.html}} routine. The routine is called 100 times and an average $b_{ij}$ is taken. Values of $a_{ij}$ are calculated so that the post-FUor area includes only post-outburst values. 

Areas of application of every relation are limited by absolute $i$ flux (on the horizontal axis of the flux-flux diagrams). The lower limit is drawn when extension of the criteria to lower fluxes stops resulting in new post-outburst points inside the post-FUor area, or when pre- and post-outburst values become hard to separate as chemistry becomes qualitatively different in discs with parameters corresponding to low fluxes. The lowest lower limit is set around the noise level (100\,mJy~km~s$^{-1}$). {Models with fluxes below the lower limit are generally the models with low disc parameters (mass and size) described in the final paragraph of Section~\ref{sec:pfflux}.} The higher limit is set roughly at the maximum flux in the sample. It must be noted that the limits are approximate as the used sample is finite and the number of models with low flux is especially small.

The results of the post-FUor area search are presented in Tables~\ref{tab:trac_no_env} and~\ref{tab:trac_env} for cases of non-embedded and embedded discs, respectively. 
Note that the tables also contain the parameters for CO isotopologue line pairs: CO/$^{13}$CO and CO/C$^{18}$O in Table~\ref{tab:trac_no_env} and  CO/C$^{18}$O in Table~\ref{tab:trac_env}. These lines are widely used as disc mass and size tracers~\citep{2011ARA&A..49...67W,2016ApJ...828...46A,2018ApJ...859...21A}. We include them in our list of promising tracers of past FUor outbursts, despite that their use could be ambiguous. The applicability of these line pairs as post-FUor tracers will be discussed further in Section~\ref{sec:COdifficult}.

\begin{table}
    \centering
    \caption{The values of relation constants for the proposed tracers for discs with no envelope. The criterion sign column shows which sign is used to replace = in Eq.~\ref{eq:ex} to get post-FUor area.}
    \begin{adjustbox}{width=0.5\textwidth,center}
    \begin{tabular}{cccccccc}
\hline
$j$ & Criterion & $i$ & $a_{ij}$ & $b_{ij}$ & $c_{ij}$ & ${p_{ij}}$ & $i$ flux range\\
 & sign &  &  &  &  &  & Jy km/s\\\hline\hline
$^{13}$CO & > & \multirow{7}{\widthof{CO}}{CO} & $-0.058$ & 0.899 & 0 & 1 & [1.87, 20.66]\\
C$^{18}$O & > &  & 0.258 & 0.655 & 0 & 1 & [1.57, 20.66]\\
CS & > &  & 0.080 & 0.847 & 0 & 1 & [1.57, 20.66]\\
HCO$^+$ & < &  & $-0.829$ & 1.202 & 0 & 1 & [1.57, 20.66]\\\cmidrule{1-2}\cmidrule{4-8}
\multirow{2}{\widthof{N$_2$H$^+$}}{N$_2$H$^+$} & < &  & $-5.095$ & 2.123 & 0 & 1 & [1.87, 15.53]\\
 & < &  & $-4.385$ & 1.954 & 0 & 1 & [15.53, 20.66]\\\hline
C$^{18}$O & > & \multirow{6}{\widthof{H$_2$CO}}{H$_2$CO} & 1.467 & 0.388 & 0 & 1 & [0.26, 8.93]\\
CS & > &  & 2.139 & 0.409 & 0 & 1 & [0.26, 8.93]\\
HCO$^+$ & < &  & 2.115 & 0.557 & 0 & 1 & [0.31, 8.93]\\\cmidrule{1-2}\cmidrule{4-8}
\multirow{2}{\widthof{N$_2$H$^+$}}{N$_2$H$^+$} & < &  & $-0.360$ & 1.163 & 0 & 1 & [0.30, 0.40]\\
 & < &  & 0.106 & 0.984 & 0 & 1 & [0.40, 8.93]\\\hline
$^{13}$CO & > & \multirow{7}{\widthof{HCN}}{HCN} & 1.242 & 0.570 & 0 & 1 & [2.00, 22.61]\\
C$^{18}$O & > &  & $-0.650$ & 0.842 & 0 & 1 & [2.00, 22.61]\\
CS & > &  &$ -0.572$ & 1.000 & 0 & 1 & [2.00, 22.61]\\
HCO$^+$ & < &  & 0.726 & 0.811 & 0 & 1 & [2.00, 22.61]\\\cmidrule{1-2}\cmidrule{4-8}
\multirow{2}{\widthof{N$_2$H$^+$}}{N$_2$H$^+$} & < &  & $-2.735$ & 1.546 & 0 & 1 & [2.00, 2.67]\\
 & < &  & $-2.350$ & 1.433 & 0 & 1 & [2.67, 22.61]\\\hline
\multirow{3}{\widthof{$^{13}$CO}}{$^{13}$CO} & > & \multirow{11}{\widthof{HCO$^+$}}{HCO$^+$} & 1.242 & 0.703 & 0.726 & 1 & [2.17, 6.15]\\
 & > &  & $-0.057$ & 0.747 & $-0.828$ & 1 & [6.15, 9.61]\\
 & > &  & 0.078 & 0.869 & 0 & 1 & [9.61, 21.80]\\\cmidrule{1-2}\cmidrule{4-8}
\multirow{2}{\widthof{C$^{18}$O}}{C$^{18}$O} & > & & 0.255 & 0.545 & $-0.828$ & 1 & [2.17, 18.87]\\
 & > &  & $-0.649$ & 1.039 & 0.726 & 1 & [18.87, 21.80]\\\cmidrule{1-2}\cmidrule{4-8}
\multirow{2}{\widthof{CS}}{CS} & > & & 2.139 & 0.733 & 2.115 & 1 & [2.17, 10.23]\\
 & > &  & $-3.068$ & 1.645 & 0 & 1 & [10.23, 21.80]\\\cmidrule{1-2}\cmidrule{4-8}
N$_2$H$^+$ & < &  & $-3.632$ & 1.767 & 0 & 1 & [2.17, 21.80]\\\hline
\multirow{2}{\widthof{$^{13}$CO}}{$^{13}$CO} & > & \multirow{6}{\widthof{N$_2$H$^+$}}{N$_2$H$^+$} & 1.242 & 0.369 & $-2.735$ & 1 & [0.10, 1.37]\\
 & > &  & 0.078 & 0.492 & $-3.632$ & 1 & [1.37, 12.74]\\\cmidrule{1-2}\cmidrule{4-8}
C$^{18}$O & > & & 0.255 & 0.335 & $-4.385$ & 1 & [0.10, 12.74]\\\cmidrule{1-2}\cmidrule{4-8}
\multirow{2}{\widthof{CS}}{CS} & > & & 2.139 & 0.351 & $-0.360$ & 1 & [0.10, 2.33]\\
 & > &  & $-3.068$ & 0.931 & $-3.632$ & 1 & [2.33, 12.74]\\\hline
    \end{tabular}
    \end{adjustbox}
    \label{tab:trac_no_env}
\end{table}

\begin{table}
    \centering
    \caption{Same as Table~\ref{tab:trac_no_env} but for embedded objects.}
    \begin{adjustbox}{width=0.5\textwidth,center}
    \begin{tabular}{cccccccc}
\hline
$j$ & Criterion & $i$ & $a_{ij}$ & $b_{ij}$ & $c_{ij}$ & ${p_{ij}}$ & $i$ flux range\\
 & sign &  &  &  &  &  & Jy km/s\\\hline\hline
C$^{18}$O & > & \multirow{2}{\widthof{CO}}{CO} & $-1.802$ & 1.303 & 0 & 1 & [1.88, 10.93]\\
H$_2$CO & > &  & $-1.744$ & 1.277 & 0 & 1 & [1.88, 10.93]\\\hline
H$_2$CO & > & CS & 0.787 & 0.684 & 0 & 1 & [0.15, 9.58]\\\hline
H$_2$CO & > & \multirow{2}{\widthof{HCN}}{HCN} & $-1.682$ & 1.257 & 0 & 1 & [1.77, 12.38]\\
N$_2$H$^+$ & < & & $-6.882$ & 2.594 & 0 & 1 & [2.66, 12.38]\\\hline
\multirow{2}{\widthof{H$_2$CO}}{H$_2$CO} & > & \multirow{4.5}{\widthof{HCO$^+$}}{HCO$^+$} & 1.038 & 2.296 & 3 & 1/3 & [1.43, 9.75]\\
 & > &  & $-1.693$ & 1.258 & 0 & 1 & [9.75, 12.96]\\\cmidrule{1-2}\cmidrule{4-8}
CS & > &  & 0.364 & 3.356 & 3 & 1/3 & [1.43, 12.96]\\
N$_2$H$^+$ & < &  & $-5.584$ & 2.252 & 0 & 1 & [2.33, 12.96]\\\hline
\multirow{2}{\widthof{H$_2$CO}}{H$_2$CO} & > & \multirow{4.5}{\widthof{N$_2$H$^+$}}{N$_2$H$^+$} & $-1.682$ & 0.485 & $-6.882$ & 1 & [0.10, 1.21]\\
 & > &  & $-1.693$ & 0.559 & $-5.584$ & 1 & [1.21, 3.43]\\\cmidrule{1-2}\cmidrule{4-8}
\multirow{2}{\widthof{CS}}{CS} & > &  & 0.364 & 2.560 & 1.173 & 1/3 & [0.10, 3.38]\\
& > &  & 0.545 & 0.914 & 0 & 1 & [3.38, 3.43]\\\hline
    \end{tabular}
    \end{adjustbox}
    \label{tab:trac_env}
\end{table}

A set of these conditions in each case (embedded and non-embedded) must essentially be treated as system because if there is an outburst in a disc, we want all line fluxes to satisfy our post-FUor criteria. So we ensure that these conditions are not mutually contradictory. We do so by finding every pair of two conditions that can be reduced to a third condition for the line pair we already have in the system. Then we compare these secondary conditions with their already present versions to construct a more robust criterion.

Essentially we find two such cases:
\begin{align}
    \begin{cases}\label{eq:fcase}
        \log_{10}{f_j} > a_{ij} + b_{ij}(\log_{10}{f_i} - c_{ij})^{p_{ij}}&\\
        \log_{10}{f_k} < a_{ik} + b_{ik}\log_{10}{f_i}&
    \end{cases}\\
\Rightarrow \log_{10}{f_j} > a_{ij} + \frac{b_{ij}}{b_{ik}^{p_{ij}}} (\log_{10}{f_k} - (a_{ik} + c_{ij}b_{ik}))^{p_{ij}}\nonumber
\end{align}
where $(i, j, k) =$ (\{CO, HCN\}, \{$^{13}$CO, C$^{18}$O, CS\}, \{HCO$^+$, N$_2$H$^+$\}); (H$_2$CO, \{C$^{18}$O, CS\}, \{HCO$^+$, N$_2$H$^+$\}) or (HCO$^+$, \{$^{13}$CO, C$^{18}$O, CS\}, N$_2$H$^+$) for disc-only models and $(i, j, k) =$ (HCN, H$_2$CO, N$_2$H$^+$) or (HCO$^+$, \{H$_2$CO, CS\}, N$_2$H$^+$) for embedded models. The second case is:

\begin{align}
    \begin{cases}\label{eq:scase}
        \log_{10}{f_j} \lessgtr a_{ij} + b_{ij}\log_{10}{f_i}&\\
        \log_{10}{f_i} \lessgtr a_{ki} + b_{ki}(\log_{10}{f_k} - c_{ki})^{p_{ki}}&
    \end{cases}\\
\Rightarrow \log_{10}{f_j} \lessgtr a_{ij} + a_{ki}b_{ij} + b_{ij}b_{ik}(\log_{10}{f_k} - c_{ki})^{p_{ki}}\nonumber
\end{align}
where $(i, j, k)$ is either (HCO$^+$, \{CO, HCN, H$_2$CO\}, N$_2$H$^+$) for disc-only or (CS, H$_2$CO, \{HCO$^+$, N$_2$H$^+$\}) for embedded. Braces mean that any of the lines inside of them can be used.
Then either of these cases is compared to $\log_{10}{f_j} \lessgtr a_{kj} + b_{kj}(\log_{10}{f_k} - c_{kj})^{p_{kj}}$ and a right part with bigger (or lesser if the sign is $<$) value at a particular $f_k$ is chosen. Tables~\ref{tab:trac_no_env} and~\ref{tab:trac_env} present the conservative estimates of the post-outburst criteria, which accounts for all line pairs. Uncorrected results are present in Figures~\ref{fig:no_env_tracers1}, \ref{fig:no_env_tracers2} and~\ref{fig:env_tracers} as green dotted lines.

\subsection{Quality of the tracers}\label{sec:qual}
It is also crucial to estimate the quality of the tracers and by quality we mean how well pre- and post-outburst values are separated on flux-flux diagrams and for how long. Based on this definition we propose the following metric:
\begin{equation}\label{eq:delta}
    \Delta_{ij} (t) = \sum_{\rm models} \log_{10}{f_j (t)} - \left( a_{ij} + b_{ij} (\log_{10}{f_i (t)} - c_{ij})^{p_{ij}} \right)
\end{equation}
In case of $j = $ HCO$^+$ and N$_2$H$^+$, right-hand side is multiplied by $-1$ as its flux drops during the outburst. {We utilise the vertical distance to the criterion line because due to the outburst, synthetic flux values move mostly along the vertical axis. Furthermore, only the values inside the applicability area are used. Essentially, this metric} shows how far the fluxes are from the separating line (Eq.~\ref{eq:ex}) on a set of models at the particular time moment. Higher values of the metric mean fluxes go deeper into post-FUor areas, therefore they are easier to separate from the pre-outburst. Negative values indicate the inability of the tracer to identify a post-FUor. Time dependence of the $\Delta_{ij} (t)$ is also a significant characteristic, as it demonstrates for how long fluxes stay in post-FUor area and how fast the relaxation is.

\begin{figure}
    \includegraphics[width=\columnwidth]{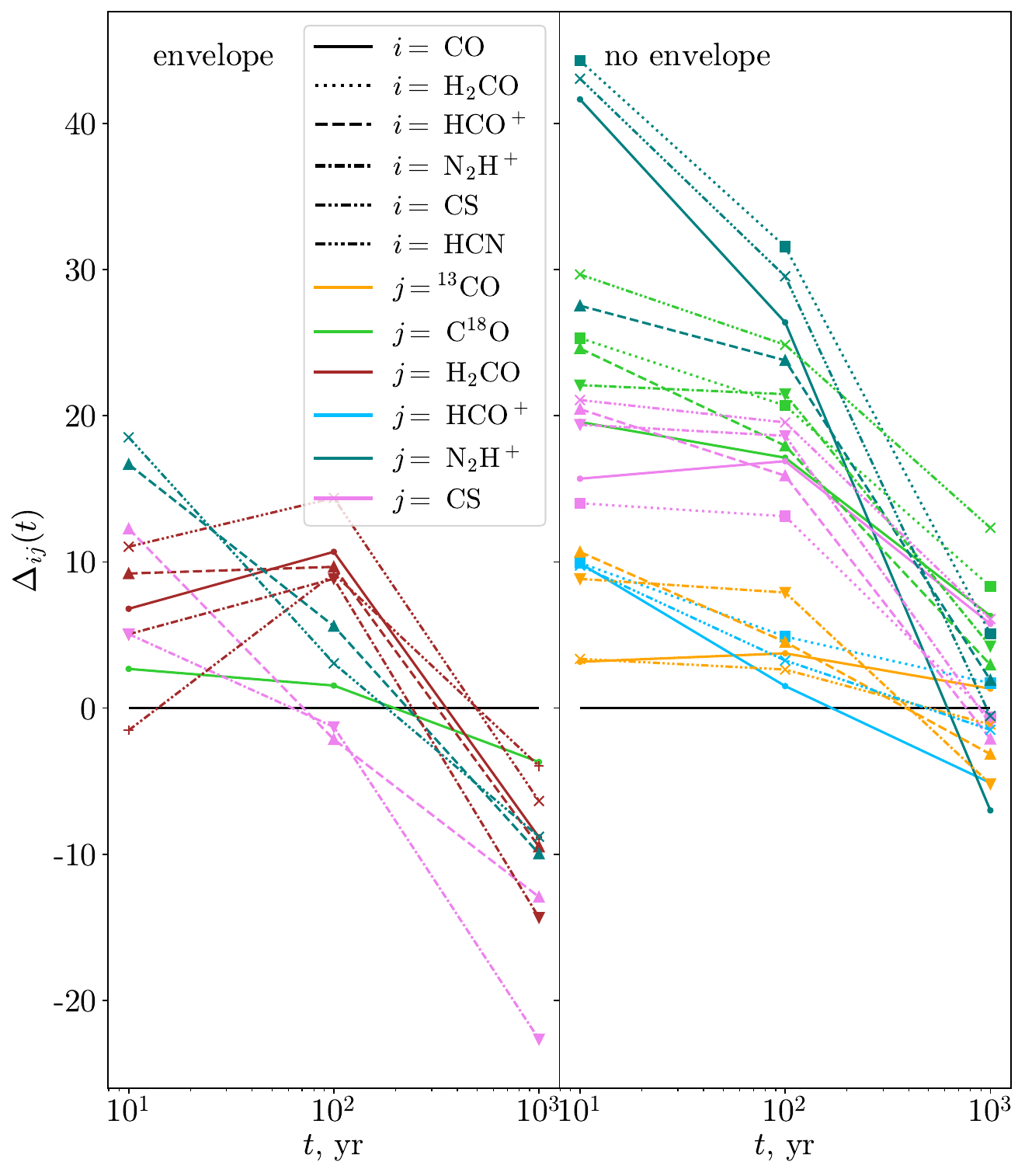}
    \caption{The values of the total distance between synthetic post-outburst fluxes and the criterion line $\Delta_{ij} (t)$  (see Eq.~\ref{eq:delta}), calculated for all line pairs identified as tracers at time moments of 10, 100 and 1000\,yr after the end of the outburst. The left panel is the case of embedded discs, the right is the case of non-embedded discs. Curve style and markers indicate the $i$ line of the pair, curve colour indicates the $j$ line of the pair.}
    \label{fig:best_tracers}
\end{figure}

Figure~\ref{fig:best_tracers} demonstrates values of $\Delta_{ij} (t)$ for all line pairs suggested as tracers at time moments of 10, 100 and 1000 yr after the end of the outburst for embedded discs (left panel) and non-embedded discs (right panel). Based on the left panel we can say that the most efficient immediate tracers are with $j = $ N$_2$H$^+$. Succeeding them are tracers with H$_2$CO which peak in efficiency at 100\,yr. Interestingly, $\Delta_{i/{\rm H}_2{\rm CO}} (100\,{\rm yr}) > \Delta_{i/{\rm H}_2{\rm CO}} (10\,{\rm yr})$, meaning that these tracers are best at finding post-FUors whose outburst passed around 100\,yr ago. The CS/H$_2$CO is special in this regard as it cannot confidently identify the very recent post-FUors. On the timescale of 1000\,yr all tracers lose their ability to find post-FUors. The varying time evolution of tracers can be used to estimate time passed since the outburst. The most convenient tracers for this are CS/H$_2$CO and HCO$^+$/CS. It can work the following way. If only HCO$^+$/CS points out a past outburst, this can indicate that it happened $\sim$10\,yr ago. If only CS/H$_2$CO tracer identifies an object as a post-FUor, then it was $\sim$100\,yr ago.
Lastly, if no tracer has the observed object in the post-FUor area, then $\sim$1000\,yr passed since the outburst. 

Focusing now on the right panel of Figure~\ref{fig:best_tracers}, we first can notice that for all tracers for discs without envelope $\Delta_{ij} (100\,{\rm yr}) > 0$. On the one hand, it means that all of them could identify a past FUor outburst around 100\,yr after its end. On the other hand, the only possible estimate of time is whether around 1000\,yr has passed since the outburst. The best tracers at 10 and 100\,yr for disc-only sources are with $j =$ N$_2$H$^+$. Next to them are the ones with $j =$ C$^{18}$O which also remain their efficiency even at 1000\,yr. Out of them the HCN/C$^{18}$O has the slowest decline and is the overall best tracer at the 1000\,yr time moment.

Figure~\ref{fig:best_tracers} also allows to make some comparison between discs with and without the envelope. The main difference is that the embedded case has less tracers, a lower max $\Delta_{ij} (t)$ and a steeper decline of $\Delta_{ij} (t)$ over all tracers. Partially this can be explained by the envelope shielding radiation coming from the disc. It dims the source and flux change seems less prominent, hence post-FUors do not go into post-FUor areas as deep or, more frequently, post-FUor areas cannot be defined at all. At the same time, longer relaxation timescales in embedded objects are explained by the difference in chemical reaction rates under different physical conditions.

\subsection{Evolution of molecular abundances and line fluxes}
\label{sec:linefluxesexplained}

There are two main reasons behind flux evolution seen in the diagrams plotted in Figures~\ref{fig:no_env_tracers1}, \ref{fig:no_env_tracers2} and~\ref{fig:env_tracers}: the effects of chemistry and radiation transfer. They are naturally interconnected, as chemical reactions can lead to such an abundance increase that the disc is no longer optically thin or, on the contrary, they can deplete the molecule revealing previously optically thick parts of the disc. 

The contribution of these two factors into flux variation is different for each molecular line. However, there are chemical processes that concern many molecules and characterise the general impact of the outburst on the disc composition. The most common effect is thermal desorption of icy species in response to the heating, which alters abundances of our target gas-phase species directly (CO, H$_2$CO, and HCN) or indirectly (HCO$^+$ and CS). Some thermally desorbed species, such as CO$_2$ and SO$_2$, are immediately involved in other gas-phase reactions. Another process concerning many species during the outburst is the enhanced photodissociation in the molecular layer at $R=10$--50\,au and at $z/R\sim0.2$. The UV~field increases during the outburst and penetrates deeper into the disc in vertical direction, affecting previously shielded molecules. This releases elemental~C (as well as elemental~O) through the photodissociation reactions \texttt{CO$_2$ $\rightarrow$ CO + O} and \texttt{CO $\rightarrow$ C + O}, and most of the carbon is immediately photoionised. The presence of C$^+$ ions triggers many chemical reactions involving carbon compounds both in the gas and on the dust surface: CO, CH, CH$_2$, CH$_4$, CH$_5^+$, C$_2$H, H$_2$CO, C$_{\rm n}$ carbon chains, as well as the formation of more complex species such as CH$_3$CHO.

Photo-dissociation in the molecular layer producing rich gas- and ice-phase chemistry is one of the key chemical processes responsible for the molecular flux changes. This effect relies on our model accounting for the adjustment of chemical structure in accordance with the density structure. Without it, during the outburst the molecular layer would stay closer to the midplane and would be unaffected by the increased radiation field. This was the case in \cite{2018ApJ...866...46M}, where despite the increased scale height the illumination of the molecular layer did not change much. The adopted assumption suggests that the disc can respond to the luminosity change fast enough, which is moderately valid for FUors (see Appendix~\ref{ap:timescales}).

In the embedded disc models, the UV radiation field in the molecular layer and at the disc periphery is higher compared to the models with no envelope due to scattering of the light on dust grains at high~$z/R$. The disc outer edge usually shielded by the body of the disc becomes additionally illuminated by the light scattered higher above the midplane. This leads to a faster dissociation of molecular species at the outer edge of the disc ($100$--$500$\,au). The effect of the outburst on the molecular layer in embedded discs turns out to be weaker. With the envelope, the radiation field is already strong before the outburst, and the area affected by the additional irradiation at higher accretion luminosities is smaller.

Among the selected molecular lines, some line fluxes after the burst are different from the ones in the quiescent state (the lines of C$^{18}$O and CS;  H$_2$CO in embedded models; $^{13}$CO and HCO$^{+}$ in non-embedded models). Others remain almost the same, despite the involvement of the corresponding species in the burst-induced chemistry (the lines of CO and HCN).
The chemical and physical processes responsible for molecular line flux changes (or lack thereof) caused by the outburst are discussed below for each of the considered molecules.

\subsubsection{CO isotopologues}
\label{sec:COchemphys}

In the majority of the models there exists a depletion area near the midplane, where most of CO is frozen out on the dust surface. During the outburst, CO in this area is sublimated from dust similarly to the results of~\cite{2017A&A...604A..15R} and~\cite{2018ApJ...866...46M}. In~\cite{2018ApJ...866...46M}, CO was not considered as a promising outburst tracer as it had less than an order of magnitude change in the abundance. So the typical timescales of elevated CO abundance (retention timescales) for {freshly} {sublimated} CO were not mentioned in~\cite{2018ApJ...866...46M}, but using the archival data from that work, we can calculate the timescales to compare with the current results. Retention timescales for CO from the modelling of~\cite{2018ApJ...866...46M} are of the order of decades. In the present research, the identically calculated retention timescales are a factor of two longer and $\sim$100\,yr{.
The total abundance change is similar, ranging from a few per cent to $\sim$3~times in~\citet{2018ApJ...866...46M} and being around 2--3 times in this study.}

\begin{figure}
    \includegraphics[width=\columnwidth]{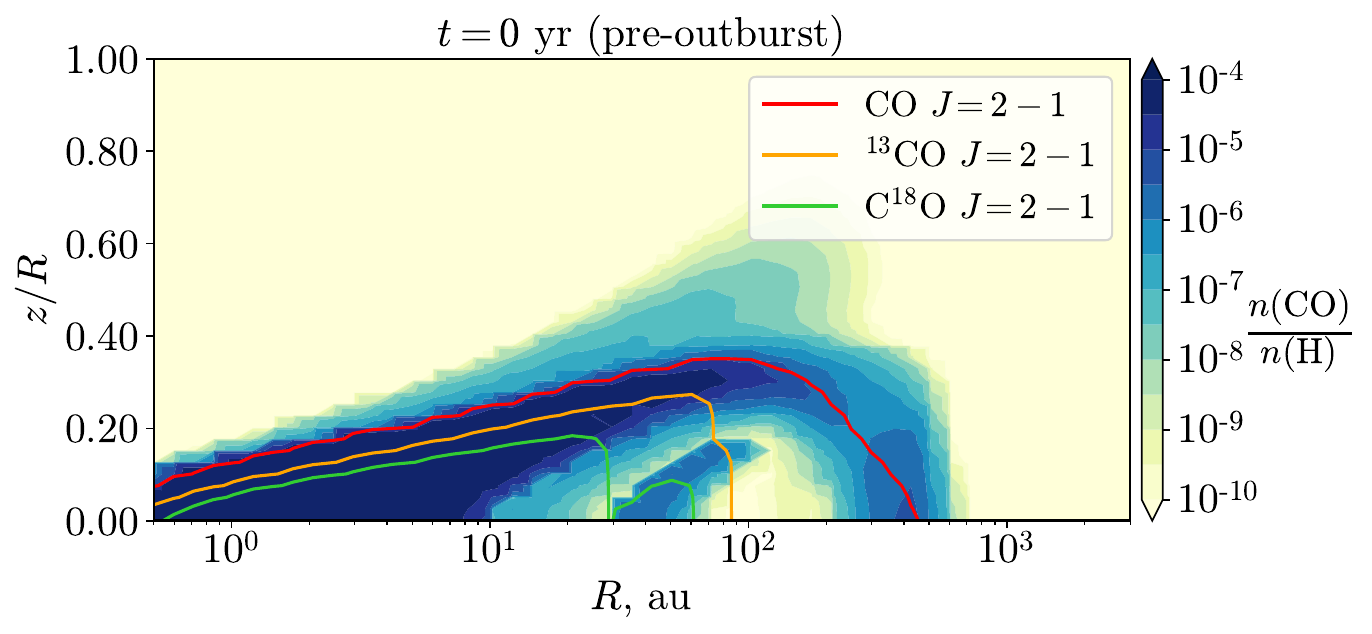}\\
    \includegraphics[width=\columnwidth]{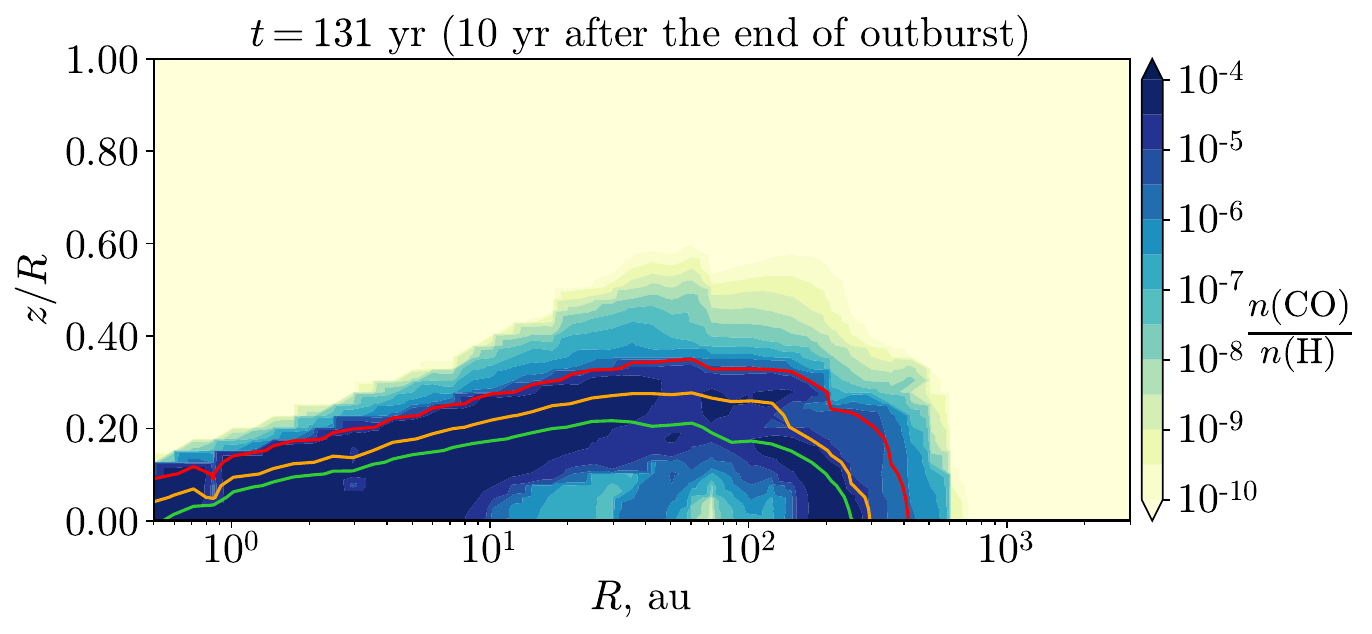}\\
    \includegraphics[width=\columnwidth]{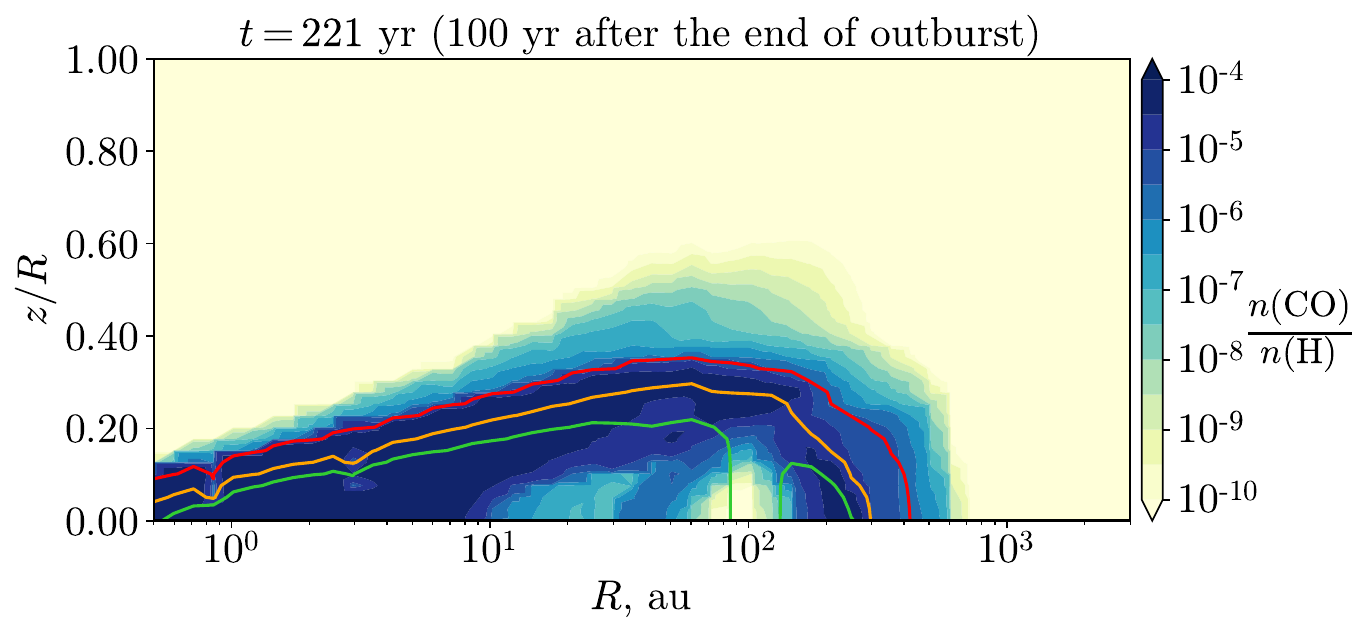}\\
    \includegraphics[width=\columnwidth]{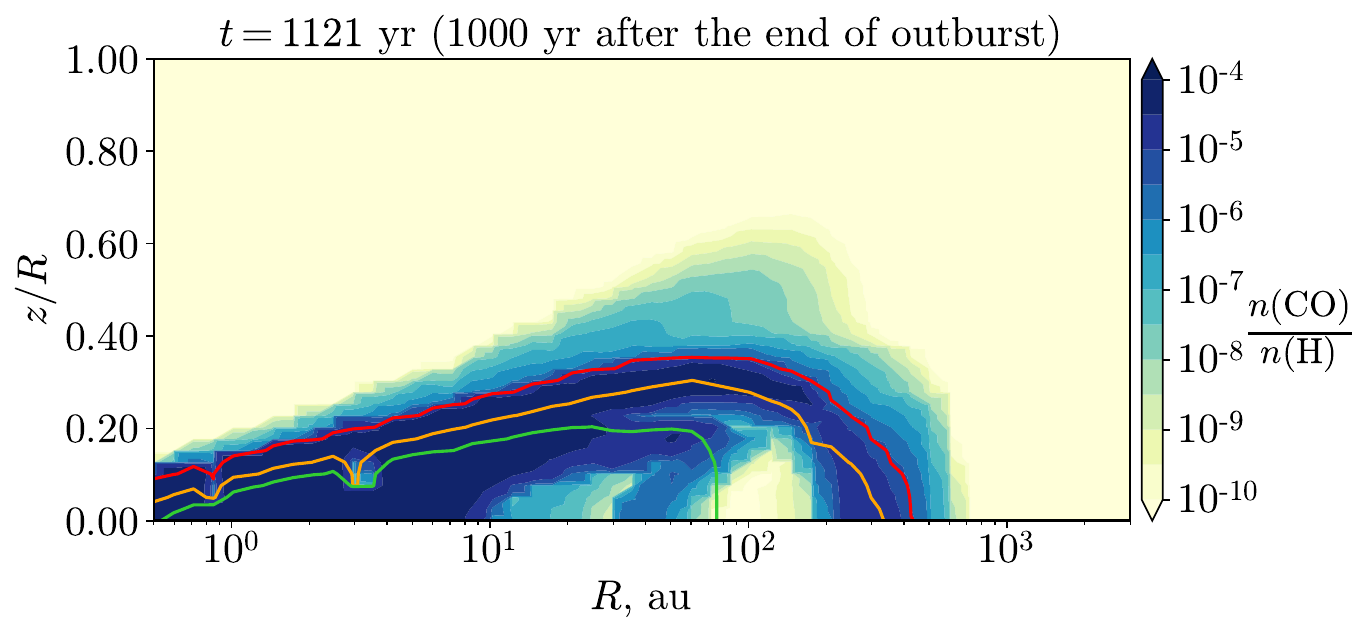}
    \caption{Abundance maps of gas-phase CO for one of the non-embedded models before the outburst and {10, 100 and 1000}\,yr after. {Model parameters are 
    $M_\star=0.8\,M_\odot$, $M_{\rm d}=0.07$\,$M_\odot$, $R_{\rm c}=50$\,au, $L_{\rm burst}=385$\,$L_\odot$.} Coloured lines are $\tau=1$ surfaces for CO isotopologue lines when looking at the disc face-on.}
    \label{fig:tau_plots}
\end{figure}

{
The difference in timescales compared to \cite{2018ApJ...866...46M} is caused by the restriction of surface reaction rates aimed to account for thick ice mantles. It slows down the conversion of CO to CO$_2$ on dust surface in the midplane and the photo-desorption of CO ice at the disc periphery. As a result, there is more CO ice by the beginning of the outburst, and the region where it is preserved is more extended toward the periphery. The re-freeze-out of CO after the outburst proceeds from inside out as can be seen from Figure~\ref{fig:tau_plots} and was previously described by \citet{2015A&A...577A.102V,2017A&A...604A..15R}. Thus the freeze-out timescales are longer in the present model, as more CO is sublimated in the outer regions with low density and temperature. Other factors, such as different binding energies and adjustment of chemical distributions to changing physical structure, have a lesser effect on CO abundance.}

However, the changes in abundance are not directly translated into changes in the line fluxes. Particularly, it is not so for typically optically thick lines, such as CO. The flux in CO~$J = 2-1$ line is affected mostly by changes in disc thermal structure, so its amplitude is limited and it returns to pre-outburst values very quickly as temperature settles down. However, for $^{13}$CO $J=2-1$ and C$^{18}$O $J=2-1$ lines the chemical impact can also be significant. This is explained by the different optical depth of these lines. Due to CO being so abundant, the $\tau = 1$ surface for this line lays in the upper disc layers close to the surface. This can be seen in abundance distributions for CO with $\tau=1$ surfaces in Figure~\ref{fig:tau_plots}. {At all times}, optical depth calculated at the centre of the CO $J = 2-1$ line equals unity (red line) above the CO ice reservoir. It means that emission from newly sublimated midplane CO does not affect the observed line flux. At the same time, $^{13}$CO $J=2-1$ and C$^{18}$O $J=2-1$ lines (orange and green in Figure~\ref{fig:tau_plots}, respectively) are optically thin in the depletion area before the outburst in non-embedded discs. In embedded discs, only the C$^{18}$O $J=2-1$ line is optically thin. After the outburst, the $^{13}$CO and C$^{18}$O (or only C$^{18}$O if there is an envelope), $\tau = 1$ surfaces move {farther from the star} to the initial CO ice reservoir, where freshly sublimated CO isotopologues emit radiation in these lines. This means that the change in flux for these lines is apparent even on 1000\,yr timescale. The CO $J = 2-1$ line, in turn, can serve a benchmark line, that is only slightly affected by the previous outburst once the temperature returns to the pre-outburst value.

The photo-dissociation of CO mentioned above is not the main contributor to either line fluxes or CO gas-phase abundance. For some period of time, its abundance even increases, as CO is the product of photodissociation of the sublimated CO$_2$. For such an abundant molecule the local increase of a factor of a few may not reflect significantly on its total abundance and line flux, but it serves as a generous source of carbon for other, less abundant molecular species.

{As was noted in the second paragraph of this Section, $^{13}$CO and C$^{18}$O (or C$^{18}$O only, if there is an envelope) the emission area expands due to the outburst. In the example presented in Figure~\ref{fig:tau_plots}, the $\tau=1$ surface for C$^{18}$O moves from 30\,au to 200\,au between 0~and 131\,yr. This opens a different way to trace post-FUors but in spatially-resolved observations. It can be done in comparison with CO emission which remains unaffected by the outburst. If C$^{18}$O- and CO-derived disc sizes appear to be similar, then an outburst might have happened recently. If not, then an outburst happened too long ago or did not happen at all. Essentially, the same method was proposed in a work by~\citet{2015A&A...579A..23J} but for envelopes. We see the same effect in discs as well.}
     
\subsubsection{H$_2$CO}  

In embedded models, there are two main reservoirs of formaldehide responsible for the change in the molecule abundance after the end of the outburst. During the outburst, H$_2$CO is thermally desorbed from dust surface in the region between the quiescent and the outburst-state snowlines (tens of au wide in the region $\sim$10--100\,au, depending on the model parameters). {In the outer disc midplane (70--100\,au) it returns to the ice several decades after the end of the outburst, the re-freeze-out proceeding inside-out.}

After the outburst, another reservoir of gas-phase H$_2$CO appears in the upper layers of the disc, in the molecular layer described above.
In this layer, most of H$_2$CO is formed after the burst from methyl radical CH$_3$ in the reaction \texttt{CH$_3$ + O}. The abundance of CH$_3$ is increased due to its formation in a chain of reactions on dust during and after the burst, triggered by the release of carbon from CO$_2$. {The newly formed H$_2$CO in the layer freezes-out at the timescales of $\sim$100\,yr.} Around $\sim$10\,yr after the outburst, H$_2$CO abundance in this upper layer is already enough for its emission to become optically thick and to partially cover the midplane reservoir. So the approximately order of magnitude rise in flux between pre-outburst and 10\,yr post-outburst moments can be attributed to both reservoirs. Later, around $\sim$100\,yr, only the upper layer reservoir is left, however the accumulated H$_2$CO is enough to increase the flux even further. Only by $\sim$1000\,yr the flux falls, as  the formed H$_2$CO freezes-out to dust.

In the absence of envelope, the main processes are the same, but occur at slightly different timescales due to density and temperature differences. Another factor is the existence of an additional reservoir of H$_2$CO  before the outburst at 300--400\,au in the midplane, at the disc periphery. There, H$_2$CO is formed in the reaction of 
\texttt{CH$_3$+O $\rightarrow$ H$_2$CO + H}, {which is only efficient until CH$_3$ is photo-dissociated. In the embedded models, this reservoir is already gone before the outburst starts, due to the increased radiation field (see Section~\ref{sec:linefluxesexplained} above). This creates an initial order of magnitude difference between pre-outburst fluxes for different envelope modes. After the outburst, this reservoir is destroyed in the model with no envelope, too. However, this does not affect much the total line flux, it remains almost the same due to the new reservoirs created by the outburst.} This small amplitude of flux change is what renders line pairs with H$_2$CO unusable as tracers in non-embedded case, unless the second line flux experiences noticeable changes.

\subsubsection{HCN} 

During the outburst, a great amount of HCN ice sublimates from dust surface at $\sim$10--200\,au. {HCN emission is already optically thick there before the outburst, so, similarly to CO, sublimation has little impact on HCN flux. Beyond this area, HCN is effectively destroyed.} Additionally, HCN forms in the gas in the molecular layer within $\sim$10\,au. {At the return to quiescent luminosity,} only a small overabundance {is left} in a thin layer at $\sim$10--100\,au. It originates both from the sublimation and gas-phase formation from H$_2$CN, which like H$_2$CO is formed from CH$_3$ in the reaction with atomic~N. This layer only slightly contributes to the HCN flux{. More significant is the destruction of HCN beyond 200\,au,} resulting in {a} small {decrease of the flux} after the outburst.
    
\subsubsection{CS}

For CS, the main flux changes are linked with the chemistry in the molecular layer. Although thermal desorption does take place, it is not substantial, because there is little ice to start with and what sublimates quickly returns back. The main formation route for this molecule is \texttt{C$_2$ + S $\rightarrow$ CS + C}, it is part of the ensemble of chemical reactions activated by photodissociation. The necessary C$_2$ comes from the interconnected chains of reactions among carbon-bearing species, starting from elemental and ionized C; the S~atoms are released from SO and H$_2$S by photodissociation and neutral-neutral reactions with the atomic hydrogen.
After the outburst, when the disc cools and settles back, CS continues to form from the excess C and S to later freeze out onto dust. For CS, this is a relaxation process, and the abundance simply returns to its pre-outburst level.

The above description of CS abundance evolution applies to both embedded and non-embedded cases. However, in the presence of an envelope, the initial radiation field is several orders of magnitude stronger due to the scattering in the envelope. Thus, the precursor molecules (CO$_2$, SO, H$_2$S) are depleted in the upper layers of the disc even before the outburst. As a result, the relaxation process is faster and of smaller amplitude. In the end, the difference between pre- and post-outburst CS flux for embedded models comes out negligible.
     
\subsubsection{HCO$^+$}
     
The HCO$^+$ destruction via sublimated neutral dipolar molecules, such as water, is a known chemical route~\citep{1998ApJ...499..777B}, which was seen in observations of protostellar envelopes~\citep{2018A&A...613A..29V} and erupting young stars~\citep{2013ApJ...779L..22J,2021A&A...646A...3L, 2017ApJ...843L...3C}. In our work we see it proceed in the disc through the reactions with H$_2$O (in the inner disc) and HCN (in outer disc) during the outburst, leading to decreased HCO$^+$ flux at the end of it. After the outburst, HCO$^+$ slowly forms again returning back to its pre-outburst abundance through reactions of \texttt{H$_3^+$ + CO} and \texttt{N$_2$H$^+$ + CO}. The N$_2$H$^+$ ion is re-formed from H$_3^+$ and H$_3^+$ is formed in reactions with simple ions already abundant in the disc.

When the envelope is present, there is little difference between pre- and post-outburst HCO$^+$ flux. It can be explained the following way. Most of the HCO$^+$ flux comes from the envelope. During the outburst, HCO$^+$ there is destroyed almost solely through recombination. Electrons for this empowered recombination are obtained by ionising hydrogen, which, in turn, is produced via the photo-destruction of H-bearing species by the increased UV radiation. When the outburst ends and the UV field returns to the pre-outburst level, HCO$^+$ replenishes through \texttt{CO$^+$ + H$_2$} reaction rather quickly at only $\sim$10\,yr timescale. 

\subsubsection{N$_2$H$^+$}

The anticorrelation between the abundances of N$_2$H$^+$ and CO is a known phenomenon~\citep{2002ApJ...570L.101B}. N$_2$H$^+$ is efficiently destroyed by the gas-phase CO, and its emission is used as a tracer of CO-poor gas both in protostellar cores~\citep{2010ApJ...711..655J} and in protoplanetary discs~\citep{2017A&A...599A.101V,2019ApJ...882..160Q}. When CO is sublimated by the outburst, N$_2$H$^+$ abundance decreases, as well as the N$_2$H$^+$ $J=3-2$ line flux. The retention timescales for this molecule are {three times lower than} those of CO{. This difference stems from N$_2$H$^+$ forming mainly in the areas where CO re-freezes out the fastest. Thus N$_2$H$^+$ timescales do follow those of CO but only locally. There is more CO mass in the colder regions, so globally CO takes more time to return to previous abundance levels.}

\section{Application to observations}\label{sec:dis}

In the following Sections we discuss the applicability of CO isotopologues to tracing past outburst (Section~\ref{sec:COdifficult}) {and} apply the proposed criteria to quiescent objects to find past FUor candidates (Section~\ref{sec:COobs}).

\subsection{Molecular lines as tracers of disc properties}
\label{sec:COdifficult}

\begin{figure*}
\includegraphics[width=2\columnwidth]{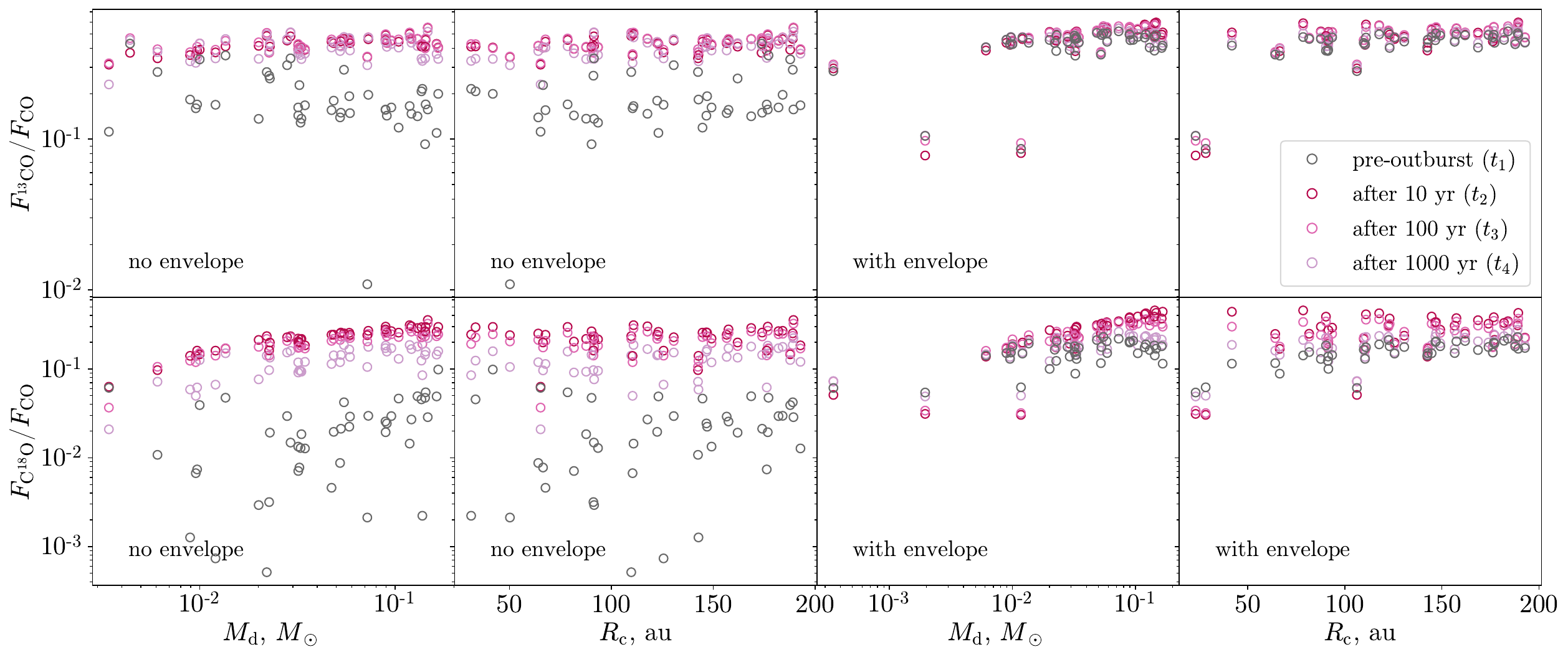}
\caption{CO/$^{13}$CO, CO/C$^{18}$O flux ratios as the functions of disc mass or characteristic disc radius for non-embedded (left panels) and embedded (right panels) objects. Colour of the markers represents time moments.}
\label{fig:lineratios}
\end{figure*}

\begin{figure}
\includegraphics[width=\columnwidth]{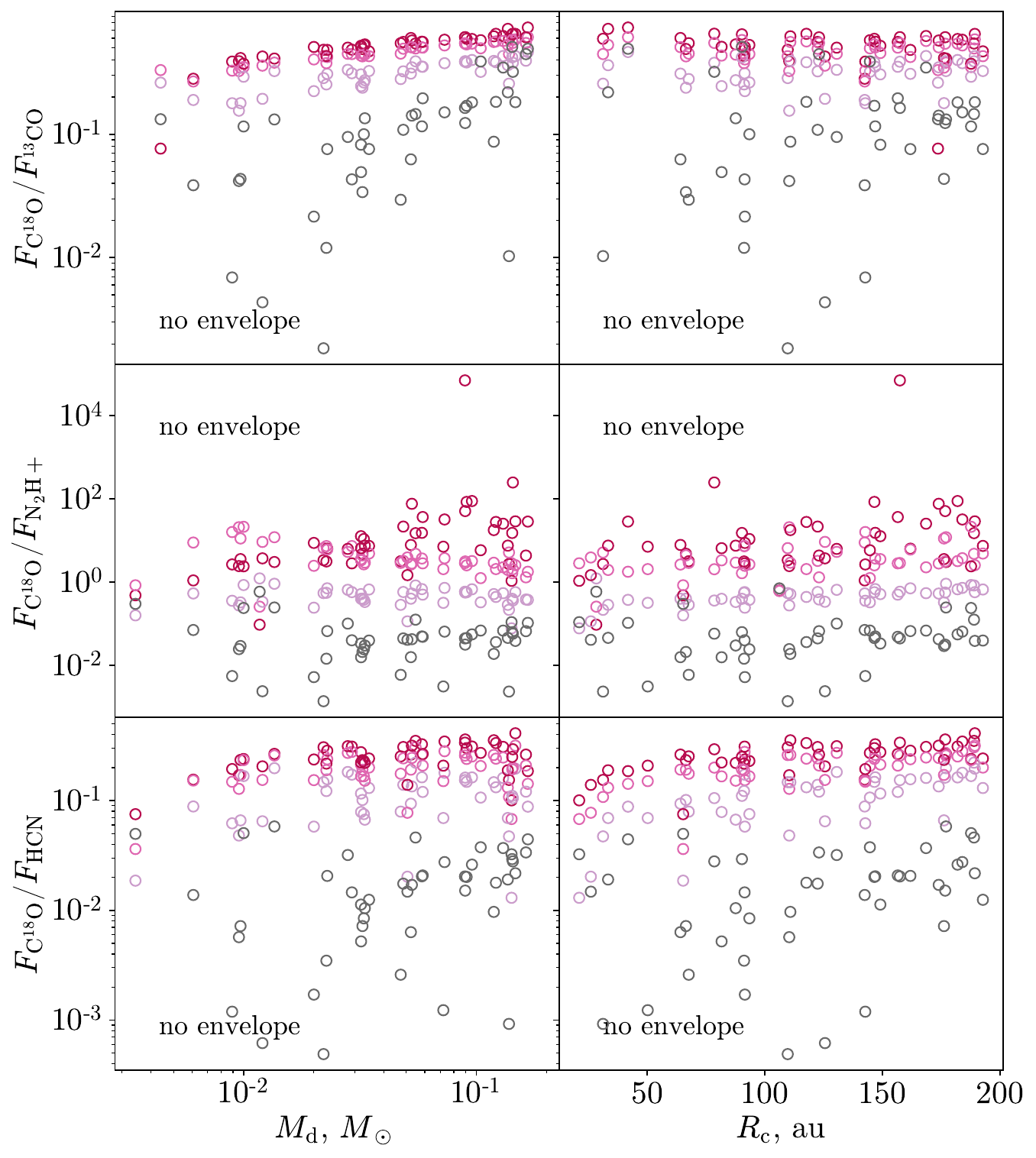}
\caption{{Same as Figure~\ref{fig:lineratios} but for $^{13}$CO/C$^{18}$O, N$_2$H$^+$/C$^{18}$O and HCN/C$^{18}$O and disc-only models.}}
\label{fig:lineratios2}
\end{figure}

The most readily observed molecular lines in protoplanetary discs are those of CO and its isotopologues~\citep{2014ApJ...788...59W}. Our modelling suggests criteria for line pairs CO/C$^{18}$O and CO/$^{13}$CO to indicate past FUor outbursts in discs with no envelope and CO/C$^{18}$O for embedded discs. However, due to its tight dependence on other system parameters, emission of CO isotopologues might seem a controversial choice for tracing past FUor outburst. It is known to trace gas mass in protoplanetary discs and the disc outer edge \citep{2016ApJ...828...46A,2018ApJ...859...21A}, and is generally used to probe gas distribution and thermal structure of the discs \citep{2003A&A...399..773D,2011ARA&A..49...67W}. In particular, the CO lines are brighter in massive discs \citep{2014ApJ...788...59W,2016A&A...594A..85M}, as well as in larger discs, regardless of the outburst history. As for the correlation between molecular emission and stellar mass, it is unclear if it exists, although disc and star masses are found to be correlated~\citep{2011ARA&A..49...67W}. As the effect of the outburst on most molecular lines is also increasing of the flux, some massive and large quiescent objects could be as bright as smaller post-outburst ones.

To solve this ambiguity, independent measurements of disc properties could be useful, most importantly, of the gas mass in the disc. Dust continuum emission provides a proxy for it \citep{2005ApJ...631.1134A}, however there are multiple sources of uncertainties, many of which arise from the evolution of dust, its optical properties and abundance relative to gas. Another way to determine gas mass, conversely, relies on dust evolution, making use of ``dust lines'' enclosing certain fraction of dust continuum emission \citep{2022A&A...657A..74F}. As kinematic data with high angular and spectral resolution become more accessible, the dynamical mass measurement for massive protoplanetary discs becomes possible \citep{2021ApJ...914L..27V,2023MNRAS.518.4481L}. Using the combination of C$^{18}$O and N$_2$H$+$ molecular lines provides a more precise measurement of gas mass in the disc, as has been shown recently by \citet{2022ApJ...926L...2T}. With the advance of independent mass measurement techniques, it will be easier to disentangle the effects of past outbursts from intrinsic disc  properties in CO isotopologue emission.

Our method of searching the past-outburst criteria intrinsically attempts to account for this possible confusion. We consider discs with a variety of masses and sizes (see Table~\ref{tab:models}), and show that they indeed produce a variety of line flux values. But it is the \textit{ratio} between the two line fluxes that changes as a result of the outburst. Larger and more massive discs are generally brighter in all CO isotopologue lines, while the outburst enhances the lines that are more optically thin. Figures~\ref{fig:lineratios} and~\ref{fig:lineratios2} present the modelled line flux ratios depending on disc mass and size. {We exclude from these Figures the models with fluxes below noise level at all times in at least one of the lines.} Unlike line fluxes, the flux ratios show weak to no dependence on disc mass and size. The outburst, on the other hand, increases the flux ratios: {from a factor of a few (CO/$^{13}$CO) to several orders of magnitude (N$_2$H$^+$/C$^{18}$O).} 
It can also be noticed that {each line flux ratio, except N$_2$H$^+$/C$^{18}$O, has a well-defined upper limit.}
This is explained by different optical depth of the lines. In non-embedded objects, CO is optically thick at all times and radii, $^{13}$CO is optically thick everywhere inside the CO snowline, and C$^{18}$O is only optically thick in smaller region inside the snowline. In embedded objects, $^{13}$CO and CO are optically thick everywhere in the disc, while C$^{18}$O is optically thick inside the snowline. There is no dependence on disc size because the fluxes scale with the size of these regions. With increasing mass, however, the region where C$^{18}$O is optically thick extends further closer to the snowline, which is reflected in the weak correlation between {ratios with C$^{18}$O} and disc mass. After the outburst, more CO is released to the gas phase, making the emission of all isotopologues optically thick at almost all radial distances inside the disc, and the flux ratios become saturated. This can be clearly seen for $F_{\rm ^{13}CO}/F_{\rm CO}$ in embedded discs where both lines are already optically thick before the outburst. {For N$_2$H$^+$/C$^{18}$O ratio limit is less defined because N$_2$H$^+$ is destroyed during the outburst, so its optical depth and flux are decreasing down to noise level.}

The new method of disc mass measurement proposed by \citet{2022ApJ...926L...2T} relies on the ratio between C$^{18}$O and N$_2$H$^+$ line fluxes to help constraining CO-based gas masses, minding the uncertainties in cosmic ray ionisation rate. In this modelling, we show that the  C$^{18}$O/N$_2$H$^+$ flux ratio is also sensitive to the luminosity outbursts (see Figure~\ref{fig:no_env_tracers2}), and can be used to trace post-outburst FUors without the envelope. {The method of~\citet{2022ApJ...926L...2T} accounts for different possible CO abundances whatever the reasons for their values may be. The outburst is one of such reasons. }

{The CO/C$^{18}$O and CO/$^{13}$CO flux ratios can also be affected by isotopologue-selective photo-dissociation of CO, which was previously shown to affect gas mass estimates of protoplanetary discs~\citep{2014A&A...567A..32M}, but is not included in our modelling. We also do not consider parametric carbon depletion, instead CO depletion from the gas is treated explicitly in the chemical reaction network, through the freeze-out and processing into CO$_2$ ice on dust grains~\citep{2017ApJ...849..130M,2018A&A...618A.182B}.}

\subsection{Searching for post-FUor objects using {the proposed criteria}}
\label{sec:COobs}

\begin{figure*}
	\includegraphics[width=\columnwidth]{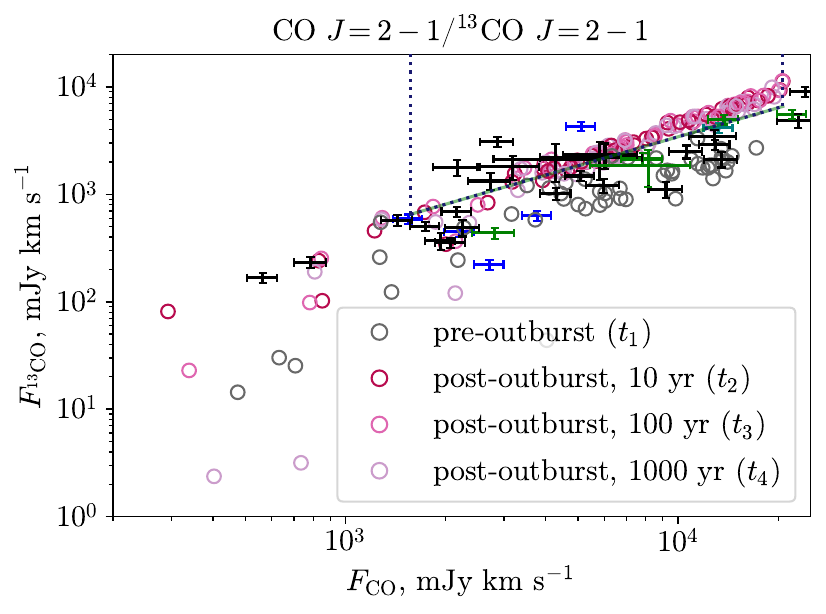}
	\includegraphics[width=\columnwidth]{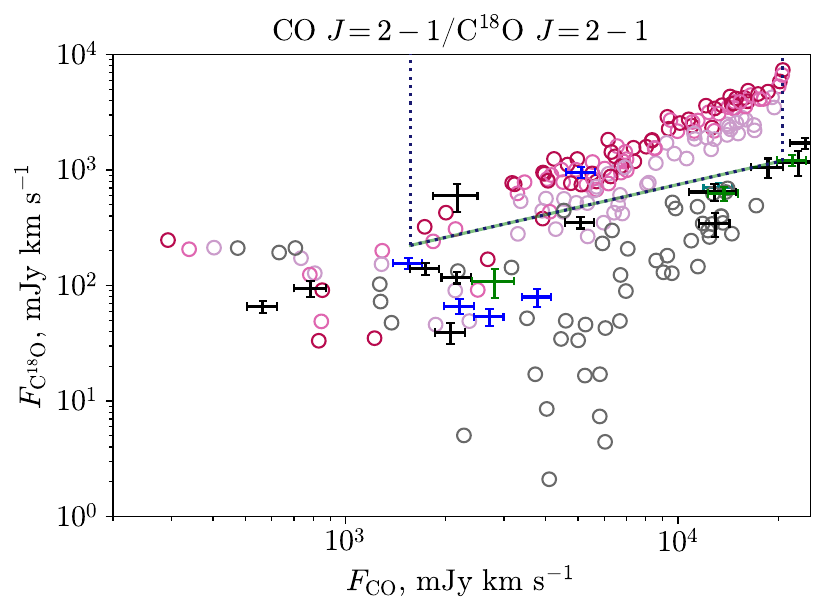}\\
    \caption{Same as Figure~\ref{fig:no_env_tracers1} (objects with no envelope) but only for CO/$^{13}$CO and CO/C$^{18}$O. Black and blue markers denote the observations of Class II discs (see Appendix~\ref{ap:obs} for the details). The blue and teal markers are from Semenov et al. (2023, submitted) and present updated data for some of the discs from the rest of the sample. The green and teal markers are observations of transitional discs (also see Appendix~\ref{ap:obs} for the details).}
\label{fig:obs_co_no_env_tracers}
\end{figure*}

Despite some complications discussed above, the most abundant data on molecular emission from protoplanetary discs is available for CO isotopologues, so we first test our criteria for CO/$^{13}$CO and CO/C$^{18}$O line pairs.
We compiled literature sources providing both CO $J=2-1$ and at least one of its isotopologue fluxes. Our sample contains 34 protoplanetary discs: all of them have CO $J=2-1$ fluxes, 33 have $^{13}$CO $J=2-1$ fluxes and 20 have C$^{18}$O $J=2-1$ fluxes (see Appendix~\ref{ap:obs}). Five discs have two separate observations. Among 34 sources, 29 are Class~II YSOs and five are transitional discs.

Flux-flux diagrams for our sample are shown in Figure~\ref{fig:obs_co_no_env_tracers}. {We} see that observational values mix rather well with synthetic ones. It can mean that our chosen tools provide an adequate description of protoplanetary discs and their emission as well as how it is received by a radio telescope.

\begin{figure*}
	\includegraphics[width=0.33\textwidth]{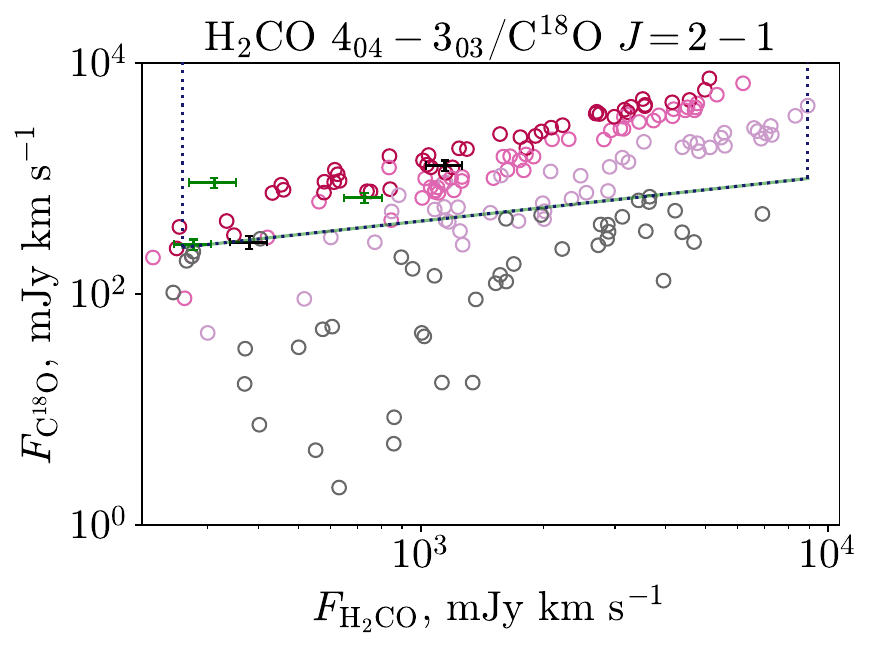}
	\includegraphics[width=0.33\textwidth]{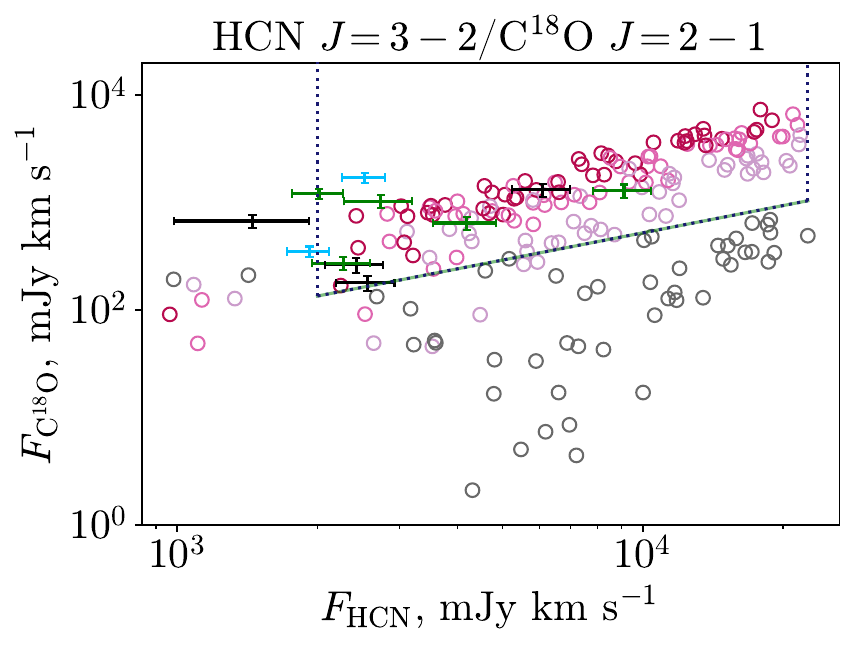}
   \includegraphics[width=0.33\textwidth]{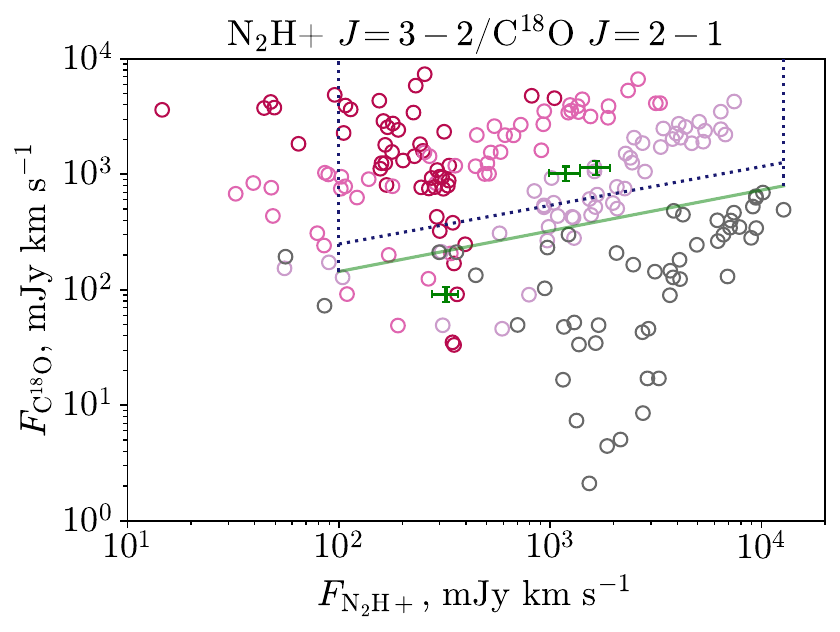}\\
    \caption{Same as Figure~\ref{fig:obs_co_no_env_tracers} (objects with no envelope) but for H$_2$CO/C$^{18}$O, HCN/C$^{18}$O {and N$_2$H$^+$/C$^{18}$O tracers. Light blue markers are observations of IM~Lup and AS~209 from~\citet{2021ApJS..257....1O}, observations of them from~\citet{2019ApJ...876...25B} are in black.}}
\label{fig:obs_hcn_no_env_tracers}
\end{figure*}

Additionally, we have found a sample of discs observed in the selected transitions (see Table~\ref{tab:lines}) of C$^{18}$O, HCN{, } H$_2$CO {and N$_2$H$^+$} lines in~\cite{2020ApJ...890..142P}{, }~\cite{2019ApJ...876...25B}{,~\cite{2021ApJS..257....1O} and~\citet{2022ApJ...926L...2T}}. Data from these papers with distances from GaiaDR3~\citep{GaiaDR3} are compiled in Appendix~\ref{ap:obs}. {Nine} discs were observed both in C$^{18}$O and HCN and five of them were observed in H$_2$CO. {Three discs were observed both in C$^{18}$O and N$_2$H$^+$.} {Six} of them are transitional discs and the other {four} are Class~II YSOs. This sample partially overlaps with our CO isotopologue sample. Namely, AS~209, CI~Tau, DM~Tau, IM~Lup, {TW~Hya, GM~Aur} 
and V4046~Sgr are present in both samples. {For IM~Lup and AS~209 we have two separate observations in HCN/C$^{18}$O.} Flux-flux diagrams for this sample are shown in Figure~\ref{fig:obs_hcn_no_env_tracers}. 

\begin{table}

\caption{{A list of post-FUor candidates from observations in several lines. Plus sign means the object fluxes are inside post-FUor area fully with their error bars. Minus sign means the object is outside the post-FUor area. Question mark means object error bars cross post-FUor area borders. More or less signs mean the object is outside higher or lower limit of criterion applicability, respectively. Zero means no data. Multiple signs refer to multiple observations. Objects with asterisks are transitional discs, those without are Class~II YSOs.}}\label{tab:trac_table}
\begin{adjustbox}{width=\columnwidth,center}
\begin{tabular}{cccccc}
\hline \hline
Object & CO/$^{13}$CO & CO/C$^{18}$O  & H$_2$CO/C$^{18}$O  & HCN/C$^{18}$O  & N$_2$H$^+$/C$^{18}$O\\\hline
{[MGM2012]} 307 & + & 0 & 0 & 0 & 0\\
{[MGM2012]} 971 & + & 0 & 0 & 0 & 0\\
{[MGM2012]} 556 & + & + & 0 & 0 & 0\\
CI Tau & ++ & + & 0 & < & 0\\
IM Lup & > & > & + & ++ & 0 \\ 
AS 209 & $-$ & $-$ & ? & +? & 0\\
DM Tau* & ?? & $--$ & + & + & +\\
LkCa 15* & 0 & 0 & + & + & 0\\
J1604-2130* & 0 & 0 & 0 & + & 0\\
GM Aur* & > & ? & 0 & ? & + \\\hline 
\end{tabular}
\end{adjustbox}
\end{table}

{A compilation of possible post-FUors found from applying the criteria is presented in Table~\ref{tab:trac_table}. The most promising post-FUor candidates are [MGM2012]~556, CI~Tau, IM~Lup and LkCa~15, based on multiple non-contradictory detections. Moreover,} IM~Lup on the {CO/$^{13}$CO and CO/C$^{18}$O} flux-flux diagrams is just outside the right border of applicability (it is brighter in CO flux than all of our models). If we are to extend the applicability of the criteria line to higher fluxes, IM~Lup will fall into the post-FUor area fully for CO/$^{13}$CO and CO/C$^{18}$O, so we can be fairly confident in this identification. AS~209, however, is just under {these} post-FUor areas. The HCN/C$^{18}$O tracer is the one with the longest timescale and is able to identify post-FUors after $>1000$\,yr (see Figure~\ref{fig:best_tracers}). The non-detection according to other tracers may be the result of the outburst taking place several kyr ago. The partial detection of DM~Tau {and GM~Aur} can be explained similarly to the AS~209's case. 

Nonetheless, this still raises a question of applicability of our criteria to transitional discs. It is unknown whether FUor-type outburst can occur in a transitional disc, however most of them accrete at rates similar to those of classical T~Tauri stars, DM~Tau being one of such examples~\citep{2022AJ....164..105F}. Other transitional discs that we found as post-FUor candidates also have factors that might have stimulated an outburst in the past. In LkCa~15 disc it is the possible presence of protoplanets~\citep{2012ApJ...745....5K, 2015Natur.527..342S}, and J1604-2130 is suggested to have a highly dynamic inner disc~\citep{2018ApJ...868...85P}. Our findings might indicate that transitional discs do undergo some type of outbursts, however another possibility is that our model cannot adequately describe such objects.

We note that two of our post-FUor candidates are Class~II discs where the presence of forming planets is suggested. There are kinematic signatures of an embedded planet in IM~Lup~\citep{2022ApJ...934L..11V} and AS~209~\citep{2023A&A...672A.125F} discs.
This might suggest that the presence of forming planets is somehow associated with outburst activity or strong impact of planets themselves on the line fluxes. {Planet-induced accretion bursts were previously suggested as a possible FUor outburst mechanism, which can be either thermal instability triggered by a planet \citep{2004MNRAS.353..841L} or tidal disruption of massive protoplanets \citep{2012MNRAS.426...70N, 2023MNRAS.523..385N}. Bow shocks from planetary embryos (even small ones) can locally heat the matter up to multiple thousand K \citep{2013ApJ...776..101B}. To  determine if this heating could have effects similar to FUor outburst, a dedicated chemical modelling of a disc containing planets is required.} The possibility and the nature of this connection should be investigated {within dedicated theoretical modelling.}

Typical time between the FUor outbursts is estimated to be of the order of 10\,kyr {for both Class I and II objects~\citep[although with a bias towards Class II objects,][]{2013MNRAS.430.2910S}. For embedded protostars, \citet{2015A&A...579A..23J} estimated the interval between burst as 20\,000\,yr.} Given that for non-embedded discs, our tracers should be sensitive to post-FUors up to $\sim$1\,kyr after their outburst (Section~\ref{sec:qual}),
probability of finding a post-FUor among quiescent YSOs is roughly 1/10. In the sample with CO isotopologue data we found 5 potential post-FUors out of 34 discs ({fraction of post-FUors} is $\sim$0.15). In the sample with {HCN, H$_2$CO, {N$_2$H$^+$,} C$^{18}$O observations,} we found {6} post-FUor candidates out of {10} discs (post-FUor fraction is {0.6}). Overall, combining both samples, we found {10} potential post-FUors out of 37 discs (post-FUor fraction is $\sim${0.27}).

The CO isotopologue sample is mostly comprised of surveys focusing on specific sky regions. It is fairly uniform with a slight bias towards brighter objects. Thus we see most objects aligning with pre-outburst models and its post-FUor fraction close to the theoretical estimate. The second sample {is rather small and consists of} individual {well-studied} discs. Therefore the sample is unrepresentative of a statistical population of YSOs. This can explain why most of the sample aligns with post-outburst models and fraction of post-FUors is as high as {0.6. In particular, the HCN/C$^{18}$O tracer identifies almost all of the tested objects as past FUors (see central panel in Figure~\ref{fig:obs_hcn_no_env_tracers}). This might be the consequence of the biased sample or the result of uncertainties in the chemical modelling of HCN, which is not well understood yet.}

{All identified post-FUor candidates are promising targets for follow-up observations in other selected transitions. Additional high spatial resolution observations in C$^{18}$O may also reveal if these discs are extended in this transition which may serve as an alternative post-FUor tracer, as suggested in Section~\ref{sec:COchemphys}.}

\section{Conclusions}\label{sec:concl}

In the present paper we studied possible observational signatures of recent past (10--1000\,yr) FUor outbursts based on emission in molecular lines. We used the ANDES astrochemical model~\citep{2013ApJ...766....8A, 2018ApJ...866...46M}, the RADMC-3D radiative transfer code~\citep{2012ascl.soft02015D} and the CASA simulator of ALMA observations~\citep{2022PASP..134k4501B} to model post-outburst protoplanetary discs both with and without envelope and calculated integrated line fluxes from a set of such discs. The molecular lines were grouped into pairs and the synthetic fluxes were plotted on flux-flux diagrams. Based on the distribution and the degree of separation of pre- and post-outburst fluxes on these diagrams, we selected several line pairs showing the largest difference between pre-outburst and post-outburst phases that can serve as tracers of past FUor outbursts. The criteria of identifying a disc as post-outburst were formulated for these tracers. The key factors responsible for the flux evolution include thermal desorption of volatiles, enhancement of gas-phase formation from the sublimated species, and optical depth effects. Finally, the tracers CO $J=2-1$/$^{13}$CO $J=2-1$, CO $J=2-1$/C$^{18}$O $J=2-1$, H$_2$CO $(J_{\rm K_a, K_c}) = 4_{04}-3_{03}$/C$^{18}$O $J = 2-1$, HCN $J = 3-2$/C$^{18}$O $J = 2-1$ {and N$_2$H$^+$ $J = 3-2$/C$^{18}$O $J = 2-1$} were applied to the observations of quiescent discs.
Our main conclusions can be summarised as follows:
\begin{enumerate}
    \item {Out of 28 line pairs we studied,} we found 10 line pairs for embedded discs and 21 line pairs for non-embedded discs showing notable difference between pre-outburst and post-outburst fluxes.
    \item The best tracers for identification of past outbursts in embedded discs are: HCN $J = 3-2$/N$_2$H$^+$ $J=3-2$ for recently ($\sim$10\,yr) ended outbursts and HCN $J = 3-2$/H$_2$CO $(J_{\rm K_a, K_c}) = 4_{04}-3_{03}$ for outbursts that have ended $\sim$100\,yr ago. None of the found tracers are able to identify past FUor outbursts after $\sim$1000\, yr. Using the combination of HCO$^+$ $J = 4-3$/CS $J = 7-6$ and CS $J = 7-6$/H$_2$CO $(J_{\rm K_a, K_c}) = 4_{04}-3_{03}$ tracers, it is possible to roughly estimate the time after the outburst.
    \item The best tracers for discs without an envelope are: H$_2$CO $(J_{\rm K_a, K_c}) = 4_{04}-3_{03}$/N$_2$H$^+$ $J=3-2$ and HCN $J = 3-2$/N$_2$H$^+$ $J=3-2$ for outbursts that have ended up to $\sim$100\,yr ago and HCN $J = 3-2$/C$^{18}$O $J=2-1$ for outbursts that ended $\sim$1000\,yr ago. All 21 tracers allow to detect a post-outburst FUor 100\,yr after the end of the outburst with some keeping this ability up to $\sim$1000\,yr.
    \item For most molecules, chemical changes after the outburst are motivated by enhanced reactions in the disc molecular layer, and are strongly connected with changes in density and radiation field.
    {\item As $^{13}$CO $J=2-1$ and C$^{18}$O $J=2-1$ fluxes are similarly altered by the outburst, mass tracing using their combination may be unreliable. Using N$_2$H$^+$ $J=3-2$ and C$^{18}$O $J=2-1$ instead per~\citet{2022ApJ...926L...2T} method is more preferable as they account for variable CO abundance.}
    \item Available observations of currently quiescent discs in $^{13}$CO $J=2-1$, C$^{18}$O $J=2-1$, CO $J=2-1${, HCN $J = 3-2$, H$_2$CO $(J_{\rm K_a, K_c}) = 4_{04}-3_{03}$ and N$_2$H$^+$ $J = 3-2$} lines {revealed ten discs that possibly experienced outburst in the past 1000\,yr. The most promising candidates are [MGM2012]~556, CI~Tau, IM~Lup and LkCa~15, based on multiple non-contradictory identifications.}
\end{enumerate}

\section*{Acknowledgements}
The work was supported by the Foundation for the Advancement of Theoretical Physics and Mathematics ``BASIS'' (20-1-2-20-1). The work of Lis Zwicky was also supported by the Ministry of science and higher education of the Russian Federation, agreement FEUZ-2023-0019. {Tamara Molyarova acknowledges support by the Ministry of Science and Higher Education of the Russian Federation (Southern Federal University, state contract GZ0110/23-10-IF). D.S. acknowledges support from the European Research Council under the Horizon 2020 Framework Program via the ERC Advanced Grant Origins 83 24 28 (PI: Th. Henning).}
This project has received funding from the European Research Council (ERC) under the European Union's Horizon 2020 research and innovation programme under grant agreement No 716155 (SACCRED).
We are also thankful to Ya.~N.~Pavlyuchenkov for his comments and advice on specific physical aspects involved in our modelling.

\section*{Data Availability}

The data that support the plots and tables within this paper and other findings of this study are available from the corresponding author upon reasonable request.



\bibliographystyle{mnras}
\bibliography{example} 




\appendix

\section{Characteristic time scales}\label{ap:timescales}

Changes in the disc's physical structure in response to the luminosity outburst are governed by cooling and heating of the matter, as well as its advective transport necessary to reach the new quasi-equilibrium state. Here we assess typical timescales of these processes in a protoplanetary disc to compare them with the typical FUor outburst duration ($\sim$100\,yr).

We assume that there is no material transport in the radial direction, and the disc is passively heated by the radiation from the central source. An important dynamic time scale is Keplerian rotation period:
\begin{equation}
    \tau_{\rm K}=\frac{2\pi R}{v_{\rm K}}=2\pi \sqrt{\frac{R^{3/2}}{GM_{\star}}}.
\end{equation}
Another useful parameter is the characteristic time of the propagation of sound waves in the vertical direction $\tau_{\rm s}$
\begin{equation}
   \tau_{\rm s}=H/c_{\rm s},
\end{equation}
where $c_{\rm s}$ is the sound speed, and $H$ is the local characteristic scale height. In ANDES, $H$ is not included explicitly, so we determine its value as $z$ at which gas density is $e^{-0.5}$ times lower than in the midplane.

The heating radiation is absorbed by gas and dust at the disc surface, in the upper layers, and for the midplane temperature $T_{\rm mp}$ to increase, this energy must reach the midplane. Radiation heating timescale $\tau_{\rm th}$ is determined both by the propagation of IR radiation of heated dust and the thermal capacity of the gas. As shown by \citet{2021ApJ...923..123W}, it can be assessed as
\begin{equation}
    \tau_{\rm th}=\frac{3}{8}\frac{c_{\rm p}\Sigma_{\rm gas}\tau_{\rm IR}}{\sigma_{\rm SB}T_{\rm mp}^3},
    \label{eq:tau_thermal}
\end{equation}
where $c_{\rm p}$ is specific heat capacity (per unit of gas mass), $\Sigma_{\rm gas}$ is local gas surface density, $\tau_{\rm IR}$ is optical depth for {disc} own thermal radiation, which corresponds to the wavelength where blackbody with $T_{\rm mp}$ has a maximum, $\sigma_{\rm SB}$ is the Stefan–Boltzmann constant. The heat can also be transported to the midplane by turbulence, then the timescales are
\begin{equation}
    \tau_{\rm turb}=\frac{H^2}{\nu}=\frac{H}{\alpha c_{\rm s}}=\frac{\tau_{\rm s}}{\alpha},
\end{equation}
where $\alpha=0.01$ is turbulent viscosity parameter~\citep{1973A&A....24..337S}.

\begin{figure}
	\includegraphics[width=\columnwidth]{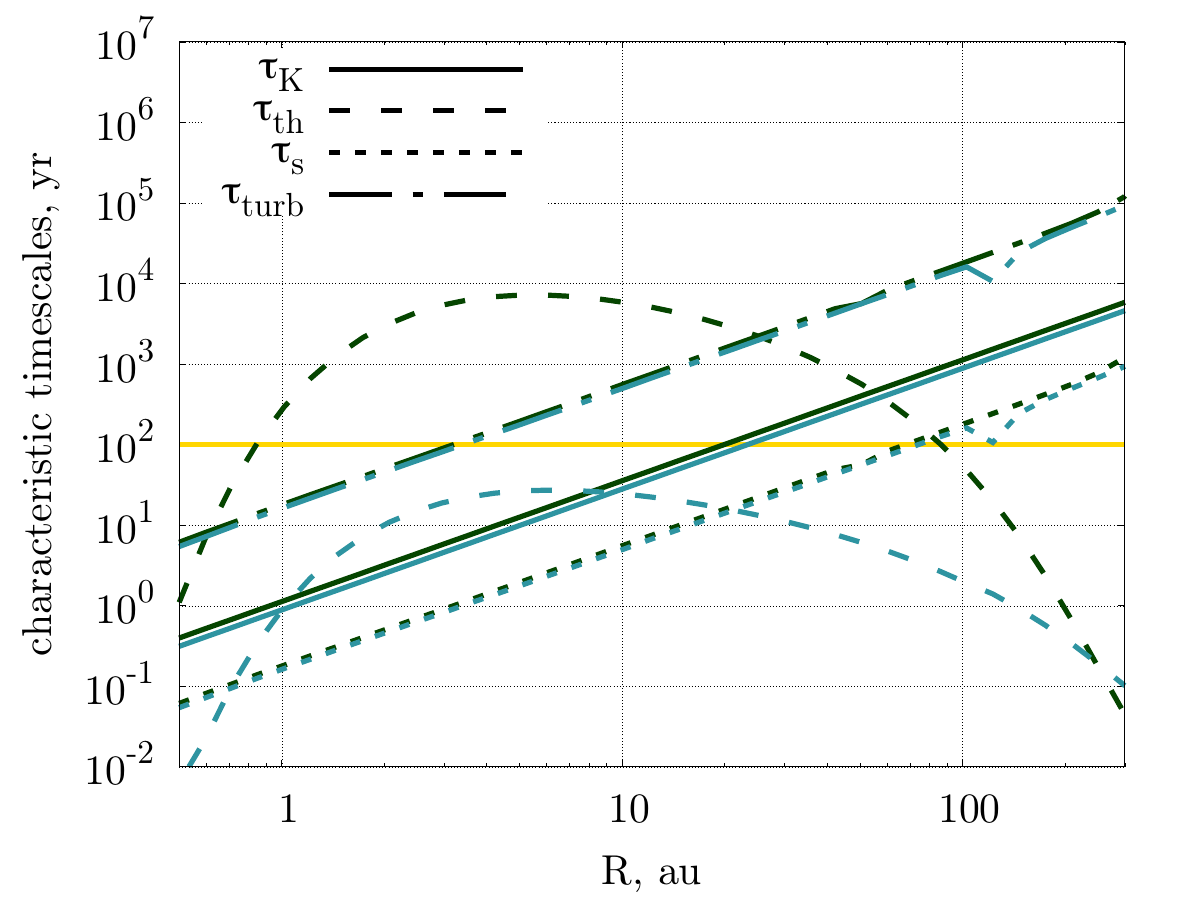}
    \caption{Characteristic timescales of physical processes in considered protoplanetary discs, depending on the distance to the star. Different shades of green correspond to two different models selected as examples. The yellow line marks the typical FUor outburst duration of 100\,yr.}
    \label{fig:timescales}
\end{figure}

These timescales for two characteristic discs of our model sample are shown in Figure~\ref{fig:timescales}. These models are selected as one with a compact and massive disc, and the other with larger and less massive disc. These are models with envelope, and the calculated timescales are only applicable to the region where the disc dominates, inside $\sim300$\,au.

For the considered low-mass stars, Keplerian timescales are below 100\,yr only inside 20--30\,au. They only give an approximate upper limit to the time needed for the density redistribution. Alternatively, gas dynamics can be characterised by $\tau_{\rm s}$, which is different from $\tau_{\rm K}$ by a factor of about $2\pi$.
These timescales indicate that outer disc regions might not reach the new equilibrium state over the outburst duration. Another lengthy process is disc radiative heating, which may take as long as tens of thousands of years. Heating is especially long for more massive discs, as $\tau_{\rm th}$ is proportional to $\Sigma_{\rm gas}^2$, while in a less massive disc ($\sim$0.03\,$M_{\odot}$) this time is below 100\,yr. This indicates that massive discs ($\sim$0.3\,$M_{\odot}$) need longer time to be heated down to the midplane by FUor outbursts. 
However, if the heating is not purely radiative, as suggested in~\citet{2021ApJ...923..123W}, but turbulent heat transport is also at play, then it can proceed somewhat faster, but still only below 100\,yr in the inner few au.

\begin{figure}
	\includegraphics[width=\columnwidth]{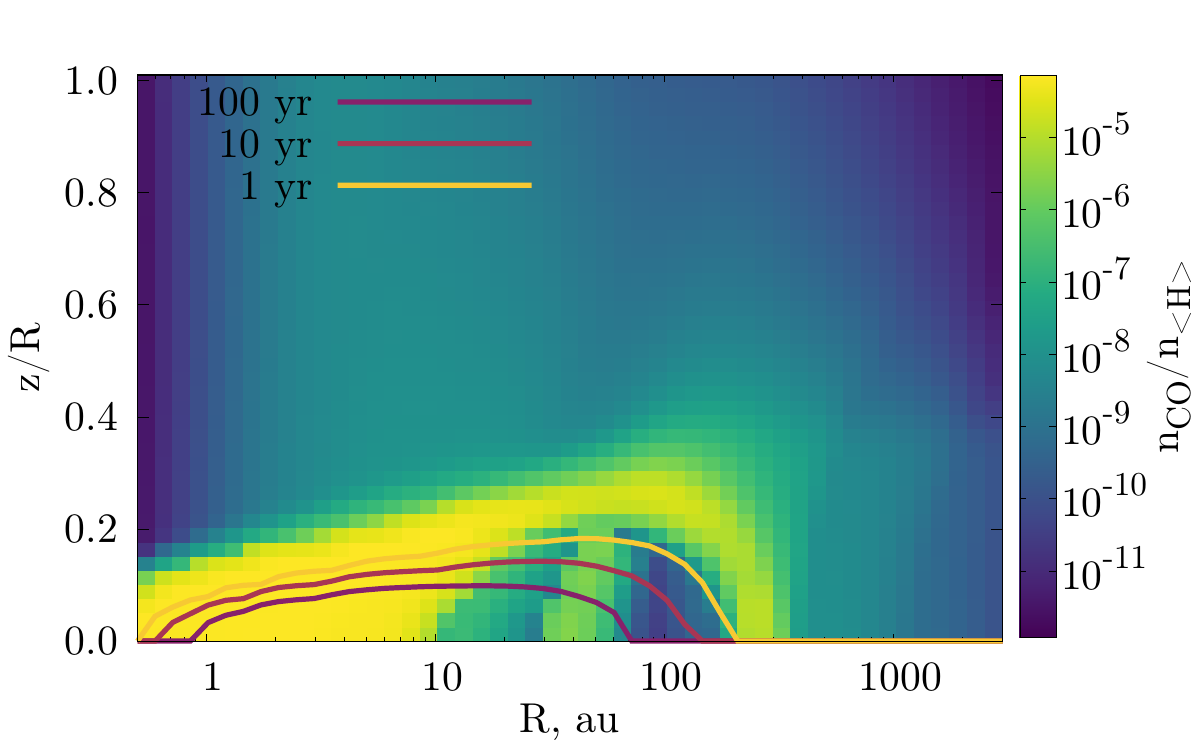}
    \caption{{Heating of the upper layers in the same model as in Figure~\ref{fig:tau_plots}. The colour shows the relative abundance of gas-phase CO before the outburst, indicating the position of the molecular layer. The lines mark the scale height above which the matter needs 1, 10 or 100\,yr to be heated by a luminosity outburst.}
    }
    \label{fig:disc_heating_time}
\end{figure}

{As seen in Figure~\ref{fig:timescales}, the duration of the burst is not enough to heat the midplane of a more massive disc. But as line emission of many molecules rather traces the molecular layer, it is also interesting to assess how deep the disc is heated during the outburst. Eq.~\eqref{eq:tau_thermal} derived by~\citet{2021ApJ...923..123W} can be modified to utilise the values of temperature, cumulative surface density and optical depth above given aspect ratio $z/R$ instead of the midplane values ($z/R=0$). Substituting these values in Eq.~\eqref{eq:tau_thermal} and assuming $t_{\rm th}=100$\,yr, for each $R$ we can get $z$ above which the matter had enough time to get heated.
In Figure~\ref{fig:disc_heating_time} we show the relative position of the molecular layer (represented by the CO spatial distribution) and the disc layers heated by the outburst. We calculate the aspect ratio to which the disc is heated from the above by the irradiation during 1, 10 and 100\,yr. The midplane between 1 and 70\,au is not heated in this model, but the molecular layer is above the critical aspect ratio even after 1\,yr. We must also note that the assessed thermal timescale is an upper limit on the heating time scale, because it relies on the (lower) pre-outburst temperatures, and is independent of the magnitude of the outburst. Turbulent heat transport mentioned above can also shorten the heating timescales.} 

The considered timescales are not negligible compared to the timescales of the luminosity change. Correct calculation of disc structure under variable illumination must be done using fully dynamic models of matter and heat transport. Our approximation is only vaguely relevant to the dynamic objects, but it allows to consider chemical processes, which are also highly dynamic. The use of a thermo-chemical model with hydrodynamics and heat transport would allow to consider the chemical effect of FUor outbursts in a more consistent manner. For EXors, with their typically shorter and less bright outbursts, this necessity is even more pressing. 

\section{Observational data}
\label{ap:obs}

This section includes tables (Tables~\ref{tab:obs_table}, \ref{tab:obs_table2} and~\ref{tab:obs_table3}) listing our compilation of observational data regarding CO isotopologues', HCN, {N$_2$H$^+$} and H$_2$CO line fluxes in low-mass protoplanetary discs that are not known to experience outbursts in the past. Flux uncertainties for all objects in both tables do not include absolute flux uncertainty and it is assumed to be 10\% of total flux. Most discs in our sample belong to Class II YSOs, except for GM~Aur~\citep{2021ApJS..257....1O}, DM~Tau~\citep{2023ApJ...948...60L}, TW~Hya~\citep{2017ApJ...837..132V}, V4046~Sgr~\citep{2015ApJ...803L..10R}, $\sigma$~Ori~540~\citep{2017AJ....153..240A}, LkCa~15~\citep{2020A&A...639A.121F} and J1604-2130~\citep{2019ESS.....450104M}, which are transitional discs.

\begin{table*}
\caption{Line fluxes for a sample protoplanetary discs with published data on CO $J=2-1$ and at least one of its isotopologues' $J=2-1$ transitions (based on a compilation of literature sources). Discs are grouped by literature references. All distances, except for [MGM2012] and $\sigma$ Ori discs, were obtained from GaiaDR3~\citep{GaiaDR3}. [MGM2012] and $\sigma$ Ori distances are from~\citet{2021ApJ...913..123G} and~\citet{2016AJ....152..213S}, respectively. Objects with an asterisk are transitional discs, those without are Class~II YSOs.}\label{tab:obs_table}
\begin{adjustbox}{width=\textwidth,center}
\begin{tabular}{ccccccc}
\hline \hline
Disc & $F_{{\rm CO}}$ & $F_{{\rm ^{13}CO}}$ & $F_{{\rm C^{18}O}}$ & Distance & $M_{\star}$ & Ref.\\
 & mJy km/s & mJy km/s & mJy km/s & pc & $M_\odot$ & \\\hline
IM Lup & $22342\pm94$ & $8370\pm83$ & $1592\pm60$ & $155.8\pm0.5$ & $\approx$1.1 & \cite{2021ApJS..257....1O}\\
GM Aur* & $19844\pm74$ & $5028\pm48$ & $1092\pm39$ & $158.1\pm1.2$ & $\approx$1.1 & \\
AS 209 & $7790\pm45$ & $2269\pm38$ & $538\pm27$ & $121.25\pm0.43$ & $\approx$1.2 & \\\hline
DM Tau* & $14900\pm400$ & $5400\pm130$ & $680\pm70$ & $144.05\pm0.46$ & $\approx$0.65 & {\cite{1997A&A...317L..55D}} \\
GG Tau & $21500\pm1600$ & $5820\pm190$ & $1080\pm100$ & $116\pm6$ & $\approx$1.2 & \\\hline
AA Tau & $5330\pm110$ & $1260\pm90$ & -- & $134.7\pm1.6$ & $\approx$0.6-0.7 & \cite{2014ApJ...788...59W} \\
CI Tau & $2510\pm130$ & $2700\pm130$ & -- & $160.3\pm0.5$ & $\approx$0.6-0.7 &  \\
CY Tau & $2020\pm80$ & $810\pm60$ & -- & $126.33\pm0.33$ & $\approx$0.6-0.7 &  \\
DL Tau & $1980\pm120$ & $430\pm60$ & -- & $159.94\pm0.5$ & $\approx$0.6-0.7 &  \\
DO Tau & $63700\pm380$ & $1520\pm100$ & -- & $138.5\pm0.7$ & $\approx$0.6-0.7 &  \\
IQ Tau & $2520\pm80$ & $480\pm70$ & -- & $131.5\pm0.6$ & $\approx$0.6-0.7 &  \\
Haro 6-13 & $17600\pm170$ & $3950\pm140$ & $470\pm100$ & $128.6\pm1.6$ & $\approx$0.6-0.7 &  \\
TW Hya* & $17500\pm1800$ & $2720\pm180$ & $680\pm180$ & $60.14\pm0.05$ & $\approx$0.6-0.7 &  \\
V4046 Sgr* & $34500\pm3500$ & $9400\pm900$ & -- & $71.48\pm0.11$ & $\approx$0.6-0.7 &  \\\hline
{[MGM2012]} 378 & $2930\pm40$ & $620\pm40$ & $150\pm30$ & $420\pm20$ & $\approx$0.4-0.5 & \cite{2021ApJ...913..123G} \\
{[MGM2012]} 556 & $290\pm30$ & $240\pm30$ & $80\pm20$ & $410\pm10$ & $\approx$0.4-0.5 &  \\
{[MGM2012]} 561 & $2290\pm30$ & -- & $130\pm20$ & $428\pm10$ & $\approx$0.4-0.5 &  \\
{[MGM2012]} 269 & $610\pm30$ & $90\pm20$ & -- & $600\pm300$ & $\approx$0.4-0.5 &  \\
{[MGM2012]} 307 & $370\pm40$ & $210\pm40$ & -- & $440\pm30$ & $\approx$0.4-0.5 &  \\
{[MGM2012]} 399 & $1320\pm60$ & $160\pm20$ & -- & $396\pm6$ & $\approx$0.4-0.5 &  \\
{[MGM2012]} 525 & $670\pm40$ & $260\pm30$ & -- & $450\pm50$ & $\approx$0.4-0.5 &  \\
{[MGM2012]} 645 & $1660\pm30$ & $260\pm30$ & -- & $428\pm10$ & $\approx$0.4-0.5 &  \\
{[MGM2012]} 762 & $1570\pm30$ &$ 370\pm30$ & -- & $390\pm10$ & $\approx$0.4-0.5 &  \\
{[MGM2012]} 971 & $410\pm40$ & $200\pm30$ & -- & $387\pm5$ & $\approx$0.4-0.5 &  \\
{[MGM2012]} 1039 & $930\pm40$ & $190\pm20$ & -- & $380\pm7$ & $\approx$0.4-0.5 &  \\\hline
$\sigma$ Ori 540* & $1204\pm85$ & $276\pm54$ & -- & $390\pm60$ & $\approx$0.79 & \cite{2017AJ....153..240A} \\
$\sigma$ Ori 1274 & $861\pm88$ & $326\pm68$ & -- & $390\pm60$ & $\approx$0.81 &  \\
$\sigma$ Ori 1152 & $633\pm82$ & $314\pm65$ & -- & $390\pm60$ & $\approx$0.62 &  \\\hline
FP Tau & $781\pm11$ & $232.0\pm8.3$ & $91.0\pm6.7$ & $127.48\pm0.42$ & $\approx$0.23 & \cite{2021ApJ...911..150P} \\
J0432+1827 & $1847\pm13$ & $537.0\pm9.4$ & $148\pm9$ & $145.6\pm1.0$ & $\approx$0.14 &  \\
J1100-7619 & $1306.0\pm6.1$ & $417.0\pm6.0$ & $71.0\pm3.6$ & $193.0\pm0.9$ & $\approx$0.23 &  \\
J1545-3417 & $738\pm10$ & $218.0\pm7.7$ & $89.0\pm9.5$ & $154.8\pm3.5$ & $\approx$0.14 &  \\
Sz 69 & $2001\pm15$ & $345\pm13$ & $38.0\pm6.6$ & $152.6\pm1.6$ & $\approx$0.2 &  \\\hline
CI Tau & $4480\pm40$ & $3750\pm40$ & $840\pm30$ & $160.3\pm0.5$ & $\approx$0.6-0.7 & Semenov et al. (2023,  \\
CY Tau & $2180\pm30$ & $830\pm20$ & $220\pm10$ & $126.33\pm0.33$ & $\approx$0.6-0.7 & submitted) \\
DL Tau & $3320\pm50$ & $560\pm20$ & $70\pm10$ & $159.94\pm0.5$ & $\approx$0.6-0.7 &  \\
DM Tau* & $14370\pm50$ & $4510\pm60$ & $760\pm20$ & $144.05\pm0.46$ & $\approx$0.65 & \\
DN Tau & $3000\pm40$ & $610\pm10$ & $90\pm10$ & $128.60\pm0.37$ & $\approx$0.65 & \\
IQ Tau & $3530\pm40$ & $290\pm10$ & $70\pm10$ & $131.5\pm0.6$ & $\approx$0.6-0.7 &  \\\hline
\end{tabular}
\end{adjustbox}
\end{table*}

\begin{table*}
\caption{Same as Table~\ref{tab:obs_table} but for a different sample with published data on C$^{18}$O $J=2-1$, HCN $J=3-2$ and H$_2$CO $(J_{\rm K_a, K_c}) = 4_{04}-3_{03}$. Distances are from GaiaDR3~\citep{GaiaDR3}.}\label{tab:obs_table2}
\begin{adjustbox}{width=\textwidth,center}
\begin{tabular}{ccccccc}
\hline \hline
Disc & $F_{{\rm C^{18}O}}$ & $F_{{\rm H_2CO}}$ & $F_{{\rm HCN}}$ & Distance & $M_{\star}$ & Ref.\\
 & mJy km/s & mJy km/s & mJy km/s & pc & $M_\odot$ & \\\hline
AS 209 & $429\pm30$ & $580\pm17$ & -- & $121.25\pm0.43$ & 0.83 & \cite{2020ApJ...890..142P}\\
DM Tau* & $998\pm17$ & $337\pm28$ & -- & $144.05\pm0.46$ & 0.53 & \\
IM Lup & $1203\pm14$ & $1063\pm25$ & -- & $155.8\pm0.5$ & 0.89 & \\
LkCa 15* & $619\pm18$ & $662\pm24$ & -- & $157.2\pm0.7$ & 1.03 & \\
V4046 Sgr* & $1184\pm18$ & $1218\pm40$ & -- & $71.48\pm0.11$ & 1.75 & \\\hline
AS 209 & $403\pm49$ & -- & $3707\pm380$ & $121.25\pm0.43$ & 0.9 & \cite{2019ApJ...876...25B}\\
CI Tau & $591\pm63$ & -- & $1271\pm390$ & $160.3\pm0.5$ & 0.8 & \\
DM Tau* & $1116\pm110$ & -- & $2966\pm400$ & $144.05\pm0.46$ & 0.5-0.6 & \\
HD 143006 & $144\pm19$ & -- & $2059\pm210$ & $167.3\pm0.5$ & 1 & \\
IM Lup & $1225\pm120$ & -- & $5654\pm570$ & $155.8\pm0.5$ & 1.0 & \\
J1604-2130* & $1377\pm140$ & -- & $9709\pm970$ & $145.3\pm0.6$ & 0.9 & \\
LkCa 15* & $589\pm63$ & -- & $3809\pm450$ & $157.2\pm0.7$ & 1.05; 0.97 & \\
V4046 Sgr* & $1203\pm120$ & -- & $10000\pm1000$ & $71.48\pm0.11$ & 0.9,0.7 & \\\hline
IM Lup & $1592\pm60$ & -- & $2343.7\pm86.3$ & $155.8\pm0.5$ & $\approx$1.1 & \cite{2021ApJS..257....1O},\\
GM Aur* & $1092\pm39$ & -- & $1814.6\pm130.5$ & $158.1\pm1.2$ & $\approx$1.1 & \cite{2021ApJS..257....6G} \\
AS 209 & $538\pm27$ & -- & $2940.8\pm71.7$ & $121.25\pm0.43$ & $\approx$1.2 & \\\hline 

\end{tabular}
\end{adjustbox}
\end{table*}

\begin{table*}
{
\caption{Same as Table~\ref{tab:obs_table} but for a different sample with published data on C$^{18}$O $J=2-1$ and N$_2$H$^+$ $J = 3-2$. Distances are from GaiaDR3~\citep{GaiaDR3}.}\label{tab:obs_table3}
\begin{adjustbox}{width=\textwidth,center}
\begin{tabular}{cccccc}
\hline \hline
Disc & $F_{{\rm C^{18}O}}$ & $F_{{\rm N_2H^+}}$ & Distance & $M_{\star}$ & Ref.\\
 & mJy km/s & mJy km/s & pc & $M_\odot$ & \\\hline
TW Hya* & $570\pm60$ & $2000\pm210$ & $60.14\pm0.05$ & 0.8 & \cite{2022ApJ...926L...2T}  \\
DM Tau* & $1110\pm110$ & $1286\pm170$ & $144.05\pm0.46$ & 0.53 & \\
GM Aur* & $1024\pm97$ & $1487\pm180$ & $158.1\pm1.2$ & 1.1 & \\\hline
\end{tabular}
\end{adjustbox}}
\end{table*}

\bsp	
\label{lastpage}
\end{document}